\documentclass[prd,preprint,superscriptaddress,amsmath,amssymb,nofootinbib]{revtex4}
\usepackage{graphicx}
\usepackage{dcolumn}
\usepackage{bm}
\usepackage{amssymb}
\usepackage{amsmath}
\usepackage{epsfig}    
\usepackage{color}
\usepackage{slashed}
\usepackage{xcolor}
\usepackage{hhline}

\def\be{\begin{equation}}
\def\ee{\end{equation}}
\newcommand{\bea}{\begin{eqnarray}}
\newcommand{\eea}{\end{eqnarray}}

\numberwithin{equation}{section}

\begin{document}

 \begin{flushright} {KIAS-P20040, APCTP Pre2021-006}  \end{flushright}

\title{Explaining Atomki anomaly and muon $g - 2$ in $U(1)_X$ extended flavour violating two Higgs doublet model} 

\author{Takaaki Nomura}
\email{nomura@kias.re.kr}
\affiliation{School of Physics, KIAS, Seoul 02455, Korea}

\author{Prasenjit Sanyal}
\email{psanyal@iitk.ac.in,prasenjit.sanyal@apctp.org}

\affiliation{Department of Physics, Indian Institute of Technology Kanpur, Kanpur 208016, India}

\affiliation{Asia Pacific Center for Theoretical Physics (APCTP) - Headquarters San 31, Hyoja-dong,
Nam-gu, Pohang 790-784, Korea}

\date{\today}

\begin{abstract}
We investigate a two Higgs doublet model with extra flavour depending $U(1)_X$ gauge symmetry where 
$Z'$ boson interactions can explain the Atomki anomaly by choosing appropriate charge assignment for the SM fermions.
For parameter region explaining the Atomki anomaly we obtain light scalar boson with $\mathcal{O}(10)$ GeV mass, 
and we explore scalar sector to search for allowed parameter space.
We then discuss anomalous magnetic dipole moment of muon and lepton flavour violating processes 
induced by Yukawa couplings of our model.
\end{abstract}
\maketitle

\section{Introduction}

The standard model (SM) of particle physics has been very successful in accommodating experimental data.
On the other hand the SM is not believed to be complete theory and existence of physics beyond the SM has been discussed.
One of the minimal extensions of the SM is introduction of new gauge symmetry such as extra $U(1)$.
Another possibility is the extension of Higgs sector introducing second Higgs doublet field so called two Higgs doublet model (THDM).
Such extensions give possibilities to explain some observed anomalies which would not be explained within the SM.

The Atomki collaboration has observed excesses of events in the Internal Pair Creation (IPC) decay of Beryllium (Be)~\cite{Krasznahorkay:2015iga, Krasznahorkay:2017qfd,Krasznahorkay:2017gwn,
Krasznahorkay:2017bwh,Krasznahorkay:2018snd} and Helium (He)~\cite{Krasznahorkay:2019lyl,Firak:2020eil} nuclei
where invariant mass and opening angle of electro-positron pair from IPC are measured showing bumps.
These excesses indicate an existence of hypothetical boson $X$ whose mass is $17.01 \pm 0.16$ MeV in $^8$Be$^*$ case and $19.68\pm0.25$ MeV in $^4$He case.
Such a light boson is expected to be induced by physics beyond the SM.
In fact there are many approaches to explain the anomaly by the new physics models; e.g. local $B-L$ model~\cite{Feng:2016jff,Feng:2016ysn,Seto:2016pks}, local Baryon number model~\cite{,Feng:2016ysn}, 
extra $U(1)$ gauge symmetry models~\cite{Gu:2016ege, Neves:2017rcn,Pulice:2019xel,DelleRose:2017xil,DelleRose:2018eic}, discussion associated with dark matter~\cite{Kitahara:2016zyb, Jia:2017iyc, Jia:2018mkc}, and flavour physics etc.~\cite{Neves:2018bay, BORDES:2019wcp,Kirpichnikov:2020tcf,Hati:2020fzp,Nam:2019osu,Wong:2020hjc,Tursunov:2020wfy,Chen:2020arr,Zhang:2020ukq,Krasnikov:2019dgh,Feng:2020mbt,Seto:2020jal}.

Muon anomalous magnetic dipole moment (muon $g-2$) is known as a long-standing anomaly
which is discrepancy between the standard model (SM) prediction and observation~\cite{PDG};
 \begin{equation}
 \Delta a_\mu = a^{\rm exp}_{\mu} -a^{\rm SM}_{\mu} = (26.8\pm 6.3 \pm 4.3)\times 10^{-10}\,.
  \label{eq:Damu}
 \end{equation}
where the deviation from the SM prediction is $\sim 3 \sigma$ level with
a positive value; recent theoretical analysis further indicates 3.7$\sigma$ deviation~\cite{Keshavarzi:2018mgv} and other analysis is also found in refs.~\cite{Aoyama:2020ynm,Davier:2019can,Davier:2017zfy,Davier:2010nc}.
Moreover, several upcoming experiments such as Fermilab E989 \cite{Grange:2015fou} and J-PARC E34 \cite{Otani:2015jra}
will give the results with more precision.
Although the recent result on the hadron vacuum polarization (HVP), calculated by  Budapest-
Marseille-Wuppertal (BMW) collaboration~\cite{Borsanyi:2020mff}, weakens the necessity of a new physics effect, it is shown in refs.~\cite{Crivellin:2020zul,deRafael:2020uif}~\footnote{The effect in modifying HVP for muon $g-2$ and electroweak precision test is previously discussed in ref.~\cite{Passera:2008jk}.} that the BMW result indicates new tensions with the HVP extracted from $e^+ e^-$ data and the global fits to the electroweak precision observables. 
If the anomaly is confirmed, the muon $g-2$ is a clear signal of a new physics effects; various solutions have been proposed to explain the anomaly such as scenarios shown in refs.~\cite{Czarnecki:2001pv,Gninenko:2001hx,Ma:2001mr,Chen:2001kn,Ma:2001md,Benbrik:2015evd,Baek:2016kud,Altmannshofer:2016oaq,Chen:2016dip,Lee:2017ekw,Chen:2017hir,Das:2017ski,Calibbi:2018rzv,Barman:2018jhz, Nomura:2016rjf,Kowalska:2017iqv,Nomura:2019btk,Nomura:2019wlo,Chen:2020jvl,Megias:2017dzd,Lindner:2016bgg,deJesus:2020upp,Liu:2020qgx,Liu:2018xkx,Terazawa:2018pdc,Kumar:2020web}. 
One of the attractive solution is given in general THDM where lepton flavour violating interactions provide sizable contribution.

In this paper we consider a model with extra $U(1)_X$ gauge symmetry and two Higgs doublets 
where the Atomki anomaly and muon $g-2$ can be explained.
The Atomki anomaly is explained by $Z'$ boson from flavour dependent $U(1)_X$ gauge symmetry introduced in ref.~\cite{Pulice:2019xel} 
which also induce relatively light scalar boson with mass of $\mathcal{O}(10)$ GeV or less from a SM singlet scalar breaking $U(1)_X$ symmetry spontaneously.
We then discuss muon $g-2$ and lepton flavour violation(LFV) processes originated from Yukawa interactions in THDM sector.
In addition neutrino mass can be generated by introducing right-handed neutrino without $U(1)_X$ charge keeping anomaly cancellation condition; 
neutrino physics is often discussed in extra $U(1)$ models~\cite{Das:2017deo,Das:2018tbd,Das:2019fee,Chiang:2019ajm,Das:2017flq,Das:2019pua,Das:2016zue,Das:2017ski,Lee:2017ekw,Nomura:2020dzw,Nomura:2020azp}. 
Taking into account light scalar boson associated with $U(1)_X$ breaking, we numerically investigate scalar potential as well as 
muon $g-2$ and LFV constraints searching for allowed parameter region.

This paper is organized as follows.
In sec.~II, we show our model and mass spectrum of new particles. In sec.~III, we discuss $Z'$ interactions explaining the Atomki anomaly.
In sec.~IV, we discuss scalar potential in the model. In sec.~V, we discuss muon $g-2$ and LFV constraints.
In sec.~VI, we briefly discuss dark sector for realizing quark mixing under flavour dependent $U(1)_X$.
Our summary is given in Sec.~VII.

\section{Model}

\begin{table}[t!]
  \centering
  \begin{tabular}{|c|c|c|c|c|c||c|c|c|} \hline
   Fields  & ~$Q_{L_i}$~ & ~$u_{R_i}$~ & ~$d_{R_i}$~ & ~$L_{L_i}$~ & ~$e_{R_i}$~ & ~$H_{1}$~ & ~$H_{2}$~ & ~$\phi$~    \\
    \hline
    $SU(3)_{C}$ & $\bm{3}$ & $\bm{3}$ & $\bm{3}$ & $\bm{1}$ & $\bm{1}$ & $\bm{1}$ & $\bm{1}$ & $\bm{1}$    \\
    \hline
    $SU(2)_L$ & $\bm{2}$ & $\bm{1}$ & $\bm{1}$ & $\bm{2}$ & $\bm{1}$ & $\bm{2}$ & $\bm{2}$ & $\bm{1}$  \\
    \hline
    $U(1)_Y$ & $\frac16$ & $\frac23$ & $-\frac13$ & $-\frac12$ & $-1$ & $\frac12$ & $\frac12$ & $0$ \\
    \hline
    $U(1)_X$ & $Q_{Q_i}$ & $Q_{u_i}$ & $Q_{d_i}$ & $Q_{L_i}$ & $Q_{e_i}$  & $Q_{H_1}$ & $Q_{H_2}$ & $Q_{\phi}$ \\
    \hline
  \end{tabular}
  \caption{Charge assignments for SM fermions under gauge symmetry including extra $U(1)_X$.   \label{tab:table1}}
\end{table}

In this section we review our model.
We consider a family non-universal $U(1)_X$ charge assignments for fermions in general two Higgs doublet model as shown in Table~\ref{tab:table1}.
If no discrete $\mathcal{Z}_2$ symmetry is imposed, both the Higgs doublets can couple to all the fermions depending on $U(1)_X$ charge assignment. 
In the flavour eigenstates of the fermions, the Yukawa Lagrangian is written by 
\begin{eqnarray}
-\mathcal{L}_\text{yukawa} &=& \bar{Q}_L Y^u_1 \tilde{H}_1 U_R  + \bar{Q}_L Y^u_2 \tilde{H}_2 U_R + \bar{Q}_L Y^d_1 H_1 D_R  + \bar{Q}_L Y^d_2 H_2 D_R \nonumber \\
&+& \bar{L}Y^l_1 H_1 e_R + \bar{L}Y^l_2 H_2 e_R + h.c,
\label{Yukawa Lagrangian}
\end{eqnarray}
where $\tilde H_{1,2} \equiv i \sigma_2 H_{1,2}^*$ with $\sigma_2$ being the second Pauli matrix.
The structure of Yukawa matrices are fixed by assignment of $U(1)_X$ charge to fermions and Higgs fields.
In the scalar sector we introduce the SM singlet $\phi$ which develops vacuum expectation value(VEV), $v_\phi$, to break $U(1)_X$ gauge symmetry.
The Higgs doublets and singlet scalar are represented as 
\begin{eqnarray}
H_i =  \begin{pmatrix}
\phi^+_i\\
\frac{v_i + h_i + i \eta_i}{\sqrt{2}}
\end{pmatrix}, \quad \phi = \frac{1}{\sqrt{2}} (\phi_R + v_\phi + i \phi_I),
\end{eqnarray} 
where $v_i$ and $v_\phi$ are the vacuum expectation values(VEVs) of the Higgs doublet $H_i$ and $\phi$ respectively, $\tan \beta = v_2/v_1$ and $v=\sqrt{v_1^2 + v_2^2}$.
We analyze scalar potential after discussing $U(1)_X$ charge assignment below.

\subsection{Constraints for $U(1)_X$ charge assignment}
Here we consider constraints for $U(1)_X$ charge assignment such as conditions to obtain the SM fermion masses and anomaly cancelation conditions associated with $U(1)_X$ and the SM gauge symmetries as discussed in ref.~\cite{Pulice:2019xel}.
To obtain fermion masses, we first impose the constraints on the charges to realize the diagonal Yukawa coupling such that 
\begin{eqnarray}
&& Q_{H_1} + Q_{d_i} - Q_{Q_i} = 0, \quad Q_{H_1} + Q_{e_i} - Q_{L_i} = 0, \quad Q_{H_1} - Q_{u_i} + Q_{Q_i} =0, \nonumber \\
&&  Q_{H_2} + Q_{d_i} - Q_{Q_i} =0, \quad Q_{H_2} - Q_{u_i} + Q_{Q_i} =0, \quad Q_{H_2} + Q_{e_i} - Q_{L_i} =0. 
\label{eq:cond_Yukawa}
\end{eqnarray}
By these conditions, we can obtain diagonal elements of Yukawa matrices and masses of the SM fermions can be realized.
We then further discuss necessary constraints for the charge assignment.

The conditions for canceling gauge anomalies are given by
\begin{align}
\label{anomaly-cond1}
 U(1)_X \times [SU(3)_c]^2 : & \ \sum_i ( 2 Q_{Q_i} - Q_{u_i} - Q_{d_i} ) =0  , \\
 \label{anomaly-cond2}
 U(1)_X \times [SU(2)_L]^2  : & \ \sum_i ( 3 Q_{Q_i} + Q_{L_i})  = 0,  \\
 \label{anomaly-cond3}
 U(1)_X \times [U(1)_Y]^2  : & \ \sum_i \left( \frac16 Q_{Q_i} - \frac13 Q_{d_i} - \frac43  Q_{u_i} + \frac12 Q_{L_i} -   Q_{e_i} \right)=0,  \\
 \label{anomaly-cond4}
  U(1)_X \times [\textrm{grav}]^2 : & \ \sum_i (  6 Q_{Q_i} -3  Q_{d_i} - 3 Q_{u_i} +2  Q_{L_i} - Q_{e_i} ) =0\\
  \label{anomaly-cond5}
 [U(1)_X]^2\times U(1)_Y : & \ \sum_i (  Q_{Q_i}^2 + Q_{d_i}^2 - 2 Q_{u_i}^2 -  Q_{L_i}^2 + Q_{e_i}^2 ) =0 \\
 \label{anomaly-cond6}
  [U(1)_X]^3 : & \ \sum_i (  6 Q_{Q_i}^3 -3  Q_{d_i}^3 - 3 Q_{u_i}^3 +2  Q_{L_i}^3 - Q_{e_i}^3 )=0.
  \end{align}
Combining the anomaly cancellation conditions and the constraints in Eq.~\eqref{eq:cond_Yukawa} we obtain 
\begin{eqnarray}
Q_{Q_1} = Q_{H_1} - Q_{Q_2} -Q_{Q_3}, \quad Q_{L_1} = -3 Q_{H_1} -Q_{L_2} -Q_{L_3}, \quad Q_{H_1} &=& Q_{H_2}, \nonumber \\
(Q_{H_2} + Q_{L_2}) (Q_{H_2} + Q_{L_3}) (2 Q_{H_2} + Q_{L_2} + Q_{L_3})=0,
\end{eqnarray}
Note that charges of two Higgs doublets are required to be identical when we require both Higgs doublets to couple with fermions satisfying anomaly cancellation conditions.
We then write $Q_{H} \equiv Q_{H_1} = Q_{H_2}$ hereafter.
Now $U(1)_X$ charges can be written in terms of $Q_{H}$, $Q_{Q_2}$, $Q_{Q_3}$ and $Q_{L_3}$ as
\begin{align}
& Q_{Q_1} = Q_H-Q_{Q_2}-Q_{Q_3}, \quad Q_{u_1} = 2Q_{H}-Q_{Q_2}-Q_{Q_3}, \quad Q_{u_2} = Q_{H}+Q_{Q_2}, \nonumber \\
& Q_{u_3} = Q_{H}+Q_{Q_3}, \quad Q_{e_1} = -2Q_H, \quad Q_{e_2} = -3Q_H-Q_{L_3}, \quad Q_{e_3} = Q_{L_3}-Q_H \nonumber \\
& Q_{L_1} = -Q_H, \quad Q_{L_2} = -2Q_H-Q_{L_3}, \quad Q_{d_1} = -Q_{Q_2}-Q_{Q_3} \nonumber \\
& Q_{d_2} = Q_{Q_2}-Q_{H}, \quad Q_{d_3} = Q_{Q_3}-Q_{H}.
\label{charge equations}
\end{align}

In our scenario we require universal charge assignment in lepton sector, which is preferred to suppress interactions between neutrinos and $Z'$ boson.
Then we impose
\begin{eqnarray}
Q_{L_3}=-Q_H
\end{eqnarray}
and this implies
\begin{equation}
Q_{L_1}=Q_{L_2}=Q_{L_3}=-Q_H, \quad Q_{e_1}=Q_{e_2}=Q_{e_3}=-2Q_H.
\end{equation}
In the universal lepton charge assignment both $H_1$ and $H_2$ can have Yukawa interaction inducing flavour changing neutral current (FCNC) interactions.
We also require the relation 
\begin{eqnarray}
Q_{Q_3}=Q_H-2Q_2.
\end{eqnarray}
The relation makes $Z'$ coupling to $d$- and $s$-quark to be the same realizing minimal flavour violation as we obtain
\begin{eqnarray}
Q_{Q_2}=Q_{Q_1}=Q_{Q_{12}}.
\end{eqnarray} 
In this case we get the other charges in terms of two independent charges $Q_{H}$ and $Q_{Q_{12}}$, given below.
\begin{align}
& Q_{Q_3} = Q_H -2Q_{Q_{12}}, \quad Q_{L_1} = Q_{L_2}=Q_{L_3}= -Q_{H}, \quad Q_{u_1} = Q_{u_2}= Q_{H} + Q_{Q_{12}} \nonumber \\
& Q_{u_3} = 2Q_{H} - 2Q_{Q_{12}}, \quad Q_{d1} = Q_{d2} = Q_{Q_{12}} - Q_{H}, \quad Q_{d3} = -2Q_{Q_{12}} \nonumber \\
& Q_{e_1} = Q_{e_2} = Q_{e_3}=-2Q_H
\label{charge equations 2}
\end{align}

Notice that in Eq.~\eqref{charge equations 2} we can assign universal charges to the quark sector by considering $Q_{Q_1}=Q_{Q_2}=Q_{Q_3}=Q_{H}/3$. 
Actually in this simple case, the charges are proportional to $U(1)_Y$ charge assignment such that 
\begin{eqnarray}
Q_{Q_{1,2,3}}=\frac{Q_H}{3}, \quad Q_{u_{1,2,3}}=\frac{4}{3}Q_H, \quad Q_{d_{1,2,3}} = -\frac{2}{3}Q_H
\label{charge equations 3}
\end{eqnarray}
where it is the same as hypercharge when we choose $Q_H =1/2$ with a different gauge coupling.
 In fact, ratio of charges are relevant and the common factor can be absorbed in normalizing $g''$.

\subsection{Gauge Sector}
Here we discuss gauge sector of the model.
The most general Lagrangian for $U(1)$ gauge sector including the kinetic mixing term is written by
\begin{equation}
\mathcal{L}_{\text{gauge}} = - \frac{1}{4}B_{\mu \nu}B^{\mu \nu} - \frac{1}{4}B^\prime_{\mu \nu}B^\prime{^{\mu \nu}} - \frac{1}{2}x B_{\mu \nu}B^{\prime \mu \nu} 
\label{Lgauge}
\end{equation}
where $B_{\mu \nu}$ and $B^\prime_{\mu\nu}$ are the  field strength tensors of $U(1)_Y$ and $U(1)_X$ gauge symmetries. Here $x(<<1)$ is a dimensionless gauge kinetic mixing parameter. 
 We can diagonalize Eq.~\eqref{Lgauge} by the following transformation
\begin{eqnarray}
\left(\begin{array}{c}
\tilde{B}^\prime_\mu\\
\tilde{B}_\mu\\
\end{array}\right)=\left(\begin{array}{cc}
\sqrt{1 - x^2} & 0 \\
x & 1 \\
\end{array}\right)\left(\begin{array}{c}
B^\prime_\mu\\
B_\mu\\
\end{array}\right)
\label{Gl2R}
\end{eqnarray}
We parameterize $\rho=-\frac{x}{\sqrt{1-x^2}}$. Under the transformation Eq.~\eqref{Gl2R}, the gauge Lagrangian can be written as 
\begin{eqnarray}
\mathcal{L}_{\text{gauge}}=-\frac{1}{4}\tilde{B}_{\mu\nu}\tilde{B}^{\mu\nu}-\frac{1}{4}\tilde{B}^\prime_{\mu\nu}\tilde{B}^{\prime\mu\nu}
\end{eqnarray}   
where $\tilde{B}_{\mu\nu}=\partial_\mu \tilde{B}_\nu - \partial_\nu \tilde{B}_\mu$ and $\tilde{B}^\prime_{\mu\nu}=\partial_\mu \tilde{B}^\prime_\nu - \partial_\nu \tilde{B}^\prime_\mu$.

The kinetic terms of the scalar fields are  
\begin{eqnarray}
\mathcal{L}_{\text{Kin}}= (D_{\mu} H_1)^\dagger(D^\mu H_1) + (D_{\mu} H_2)^\dagger(D^\mu H_2) + (D_{\mu} \phi)^\dagger(D^\mu \phi) 
\end{eqnarray}
where,
\begin{eqnarray}
D_{\mu} H_{1,2} &=& \left( \partial_\mu +igW_\mu^a\frac{\tau^a}{2}+i\frac{g^\prime}{2}\tilde{B}_\mu+i \left(\rho \frac{g^\prime}{2} - g^{\prime\prime} Q_{H}\frac{\rho}{x} \right)\tilde{B^\prime}_\mu  \right) H_{1,2} \nonumber \\
D_{\mu} \phi &=& \left( \partial_\mu - ig^{\prime\prime}Q_\phi\frac{\rho}{x}\tilde{B^\prime}_\mu \right) \phi
\end{eqnarray}
After scalar fields developing their VEVs we obtain mass terms for gauge fields, and 
those of neutral gauge fields $(W^3_\mu , \tilde{B}_\mu , \tilde{B}'_\mu)$ are 
\begin{eqnarray}
\mathcal{L}^{\text{mass}}_{\text{gauge}}= \frac{1}{2}
\left(\begin{array}{c}
W_\mu^3 \\
\tilde{B}_\mu \\
\tilde{B}^\prime_\mu \\
\end{array}\right)^T
M^2_{\text{gauge}} 
\left(\begin{array}{c}
W_\mu^3 \\
\tilde{B}_\mu \\
\tilde{B}^\prime_\mu \\
\end{array}\right).
\end{eqnarray}
The mass matrix $M^2_{\text{gauge}}$ is written by
\begin{eqnarray}
M^2_{\text{gauge}} = \frac14 \left(\begin{array}{ccc}
g^2 v^2 & -g g^\prime v^2 & gv^2(2g^{\prime\prime}Q_H -g^\prime x)\frac{\rho}{x}\\
-g g^\prime v^2 & g^{\prime 2} v^2 & g^{\prime} v^2(-2g^{\prime\prime}Q_H + g^\prime x)\frac{\rho}{x}\\
gv^2(2g^{\prime\prime}Q_H -g^\prime x)\frac{\rho}{x} & g^{\prime} v^2(-2g^{\prime\prime}Q_H + g^\prime x)\frac{\rho}{x} & 4 M_{Z^\prime}^2
\end{array}\right)
\label{mass matrix gauge boson}
\end{eqnarray}
where we parameterize 
\begin{equation}
M_{Z^\prime}^2 = \frac{1}{4}(4g^{\prime \prime 2}(Q_H^2 v^2 + Q_\phi^2 v_\phi^2)- 4g^\prime g^{\prime \prime} Q_H v^2 x + g^{\prime 2}v^2 x^2)\frac{\rho^2}{x^2}.
\end{equation}
Here we rotate $(W^3, \tilde{B})$ by Weinberg angle such that
\begin{eqnarray}
\left(\begin{array}{c}
W^3_\mu \\
\tilde{B}_\mu \\
\end{array}\right) =
\left(\begin{array}{cc}
\cos\theta_w & \sin\theta_w \\
-\sin\theta_w & \cos\theta_w\\
\end{array}\right)
\left(\begin{array}{c}
\tilde{Z}_\mu \\
\tilde{A}_\mu \\
\end{array}\right).
\end{eqnarray}
Then we obtain massless photon field as in the SM and the mass matrix for massive components is
\begin{eqnarray}
&& \mathcal{L}^{\text{mass}}_{\text{gauge}} = \frac{1}{2}
\left(\begin{array}{c}
\tilde{Z}_\mu \\
\tilde{B}^\prime_\mu \\
\end{array}\right)^T
\left(\begin{array}{cc}
M_{Z,SM}^2 & \Delta^2 \\
\Delta^2 & M_{Z^\prime}^2
\end{array}\right)
\left(\begin{array}{c}
\tilde{Z}_\mu \\
\tilde{B}^\prime_\mu \\
\end{array}\right)  \nonumber \\
&& \Delta^2 = \Big(\frac{2g^{\prime\prime}Q_H - g^\prime x}{4}\Big)\frac{\rho}{x}v^2\sqrt{g^2 + g^{\prime 2}}, \quad M_{Z,SM}^2 = \frac{1}{4}v^2(g^2 + g^{\prime 2}).
\label{masss matrix gauge boson 2}
\end{eqnarray}
The physical masses of the neutral gauge bosons are 
\begin{eqnarray}
m_Z^2 &=& \frac{1}{2}\Big[M_{Z,SM}^2 + M_{Z^\prime}^2 + \sqrt{(M_{Z,SM}^2 - M_{Z^\prime}^2)^2 + 4\Delta^4} \Big] \nonumber \\
m_{Z^{\prime}}^2 &=& \frac{1}{2}\Big[M_{Z,SM}^2 + M_{Z^\prime}^2 - \sqrt{(M_{Z,SM}^2 - M_{Z^\prime}^2)^2 + 4\Delta^4} \Big].
\end{eqnarray} 
The mass matrix in Eq.~\eqref{masss matrix gauge boson 2} can be diagonalized by rotation matrix 
\begin{eqnarray}
&& \left(\begin{array}{c}
\tilde{Z}_\mu\\
\tilde{B}^\prime\mu\\
\end{array}\right)=
\left(\begin{array}{cc}
\cos\theta & -\sin\theta\\
\sin\theta & \cos\theta \\
\end{array}\right)
\left(\begin{array}{c}
Z_\mu\\
Z^\prime\mu\\
\end{array}\right) \\
&& \tan 2\theta = \frac{2\Delta^2}{M_{Z,SM}^2 - M_{Z^\prime}^2}
\end{eqnarray}
where $Z_\mu$ and $Z'_\mu$ are mass eigenstates for the SM $Z$ and extra $Z'$ bosons. 
In summary, the original gauge fields are transformed into mass eigenstates by
\begin{eqnarray}
&& \left(\begin{array}{c}
W_\mu^3 \\
B_\mu \\
B^\prime _\mu \\
\end{array}\right) =
\left(\begin{array}{ccc}
R^w_{11} & R^w_{12} & R^w_{13}\\
R^w_{21} + R^w_{31}\rho & R^w_{22} + R^w_{32}\rho & R^w_{23} + R^w_{33}\rho \\
-R^w_{31}\frac{\rho}{x} & -R^w_{32}\frac{\rho}{x} & -R^w_{33}\frac{\rho}{x}\\ 
\end{array}\right)
\left(\begin{array}{c}
Z_\mu \\
A_\mu \\
Z^\prime_\mu \\
\end{array} \right) 
\label{eq:gauge field}
\\
&& R^w_{ab}=\begin{pmatrix}
 \cos{\theta_w}\cos{\theta} & \sin{\theta_w} & -\cos{\theta_w}\sin{\theta} \\
-\cos{\theta}\sin{\theta_w} & \cos{\theta_w} & \sin{\theta_w}\sin{\theta} \\
\sin{\theta}& 0 &\cos{\theta}
\end{pmatrix}_{ab},
\label{Rw matrix}
\end{eqnarray}
where $a(b) = 1,2,3$.

\subsection{Scalar sector}

Here we discuss scalar sector under our $U(1)_X$ charge assignment formulating mass spectrum and corresponding mass eigenstate.
The scalar potential in the model is written as 
\begin{eqnarray}
V &=& \ m^2_{H_1} H_1^\dagger H_1 + m^2_{H_2} H_2^\dagger H_2 - m^2_{H_{12}}( H_1^\dagger H_2 + h.c.) + m^2_{\phi} \phi^* \phi  + \lambda_{1} (H_1^\dagger H_1)^2 + \lambda_{2} (H_2^\dagger H_2)^2 \nonumber \\
 &+& \lambda_3 (H_1^\dagger H_1)(H_2^\dagger H_2)+ \lambda_4 (H_1^\dagger H_2)(H_2^\dagger H_1) + \frac{\lambda_{5}}{2} \{  (H_1^\dagger H_2)^2 + h.c. \} + \lambda_{6} \{(H_1^\dagger H_1)(H_1^\dagger H_2) + h.c. \} \nonumber \\
  &+& \lambda_{7} \{(H_2^\dagger H_2)(H_1^\dagger H_2) + h.c. \} + \lambda_{8} (H_1^\dagger H_1)(\phi^* \phi) + \lambda_{9} (H_2^\dagger H_2)(\phi^* \phi) +  \lambda_{10} \{ (H_1^\dagger H_2)(\phi^* \phi) + h.c. \} \nonumber \\
   &+& \lambda_{11} (\phi^* \phi)^2,
\end{eqnarray}
where we choose all the couplings to be real for simplicity.
The VEVs can be obtained by solving the conditions $\partial V/\partial v_1 = \partial V/\partial v_2 = \partial V/\partial v_\phi =0$ which require the parameters to satisfy
\begin{align}
& m_{H_1}^2v_1 - m_{H_{12}}^2 v_2  + \frac{v_1}{2} (2 v_1^2 \lambda_1 + v_2^2 \bar \lambda + v_\phi^2 \lambda_8 )=0 \nonumber \\
& m_{H_2}^2v_2 - m_{H_{12}}^2 v_1 + \frac{v_2}{2} (2 v_2^2 \lambda_2 + v_1^2 \bar \lambda + v_\phi^2 \lambda_9 )=0 \nonumber \\ 
& v_\phi (2 m_\phi^2 + 2 v_\phi^2 \lambda_{11} + v_1^2 \lambda_8 + v_2^2 \lambda_9)=0,
\end{align}
where $\bar \lambda = \lambda_3 + \lambda_4 + \lambda_5$ and for simplicity we consider $\lambda_6,\lambda_7$ and $\lambda_{10}$ to be zero.

The mass eigenstates for charged components are obtained as
\begin{equation}
\left( \begin{array}{c} G^\pm \\ H^\pm \end{array} \right)
= \left( \begin{array}{cc} \cos \beta & - \sin \beta \\ \sin \beta & \cos \beta \end{array} \right) 
\left( \begin{array}{c} \phi_1^\pm \\ \phi_2^\pm \end{array} \right),
\label{Eq:chargedMES}
\end{equation}
where $\tan \beta = v_2/v_1$, $G^\pm$ is Nambu-Goldstone(NG) boson absorbed by $W^\pm$ and $H^\pm$ is physical charged Higgs boson. 
The mass of charged Higgs boson is given by
\begin{equation}
m_{H^\pm}^2 =  \frac{m_{H_{12}}^2}{\sin \beta \cos \beta} - \frac{v^2}{2} ( \lambda_4 + \lambda_5), 
\end{equation}
where $v = \sqrt{v_1^2 + v_2^2}$.

The mass eigenstates for CP-odd neutral scalar bosons can be given by
\begin{eqnarray}
\begin{pmatrix}
\eta_1 \\ \eta_2 \\ \phi_I\\
\end{pmatrix}=
\left(\begin{array}{ccc}
\cos\beta & -\sin\beta & 0\\
\sin\beta & \cos\beta & 0 \\
0 & 0 & 1\\ 
\end{array}\right)
\begin{pmatrix}
G^0_1\\A^0\\ G^0_2\\
\end{pmatrix},
\end{eqnarray}
where $G^0_1$ and $G^0_2$ are two goldstone modes absorbed by $Z$ and $Z'$ bosons, and $A^0$ is massive CP-odd scalar. 
Then mass of $A^0$ is given by
\begin{equation}
m_{A^0}^2 = \frac{m_{H_{12}}^2}{\sin \beta \cos \beta} - \lambda_5 v^2.
\end{equation}

The CP-even scalar sector has three physical degrees of freedom $\{h_1, h_2, \phi_R \}$ and the mass matrix is given by
\begin{equation}
\mathcal{L} \supset \frac{1}{2}
\left( \begin{array}{c} h_1 \\ h_2 \\ \phi_R \end{array} \right)^T 
\left(
\begin{array}{ccc}
 \frac{v_1 m_{H_{12}}^2}{v_2} + 2 \lambda_1 v_1^2 & - m_{H_{12}}^2 \bar \lambda v_1 v_2 & \lambda_8 v_1 v_\phi \\
 - m_{H_{12}}^2 \bar \lambda v_1 v_2 &  \frac{v_2 m_{H_{12}}^2}{v_1} + 2 \lambda_2 v_2^2 & \lambda_9 v_2 v_\phi  \\
 \lambda_8 v_1 v_\phi & \lambda_9 v_2 v_\phi & 2 \lambda_{11} v_\phi^2 \\
\end{array}
\right) 
\left( \begin{array}{c} h_1 \\ h_2 \\ \phi_R \end{array} \right).
\end{equation}
The mass matrix can be diagonalized by an orthogonal matrix $R$ with three Euler parameters $\{ \alpha_1, \alpha_2, \alpha_3 \}$ which is written as
\begin{equation}
\\ R(\alpha_1,\alpha_2,\alpha_3)=\left(\begin{array}{ccc}
c_{\alpha_1}c_{\alpha_2} & - s_{\alpha_1}c_{\alpha_2} & s_{\alpha_2} \\
- c_{\alpha_1}s_{\alpha_2}s_{\alpha_3} + s_{\alpha_1}c_{\alpha_3} & c_{\alpha_1}c_{\alpha_3}+s_{\alpha_1}s_{\alpha_2}s_{\alpha_3} & c_{\alpha_2}s_{\alpha_3}\\
-c_{\alpha_1}s_{\alpha_2}c_{\alpha_3}-s_{\alpha_1}s_{\alpha_3} & - c_{\alpha_1}s_{\alpha_3} + s_{\alpha_1}s_{\alpha_2}c_{\alpha_3}& c_{\alpha_2}c_{\alpha_3}\\
\end{array}\right)
\end{equation} 
and mass eigenstates are obtained such that
\begin{equation}
\left( \begin{array}{c} h_1 \\ h_2 \\ \phi_R \end{array} \right) = R_{ij} \left( \begin{array}{c} H^0 \\ h^0 \\ \xi^0 \end{array} \right)_j.
\label{Eq:CP-even}
\end{equation}

\subsection{Quark masses and Yukawa interactions}

Here we consider Yukawa couplings for quarks 
\begin{equation}
-\mathcal{L}^{q}_{\rm Y}=(Y^u_1)_{ij}\bar{Q}_{iL}\tilde{H}_1u_{jR} + (Y^u_2)_{ij}\bar{Q}_{iL}\tilde{H}_2u_{jR} + (Y^d_1)_{ij}\bar{Q}_{iL}H_1d_{jR} + (Y^d_2)_{ij}\bar{Q}_{iL}H_2d_{jR} + h.c.  
\end{equation}
Under our $U(1)_X$ charge assignment structure of Yukawa matrices are 
\begin{equation}
Y^{u,d}_{1,2} = \begin{pmatrix} \times & \times & 0 \\ \times & \times & 0 \\ 0 & 0 & \times \end{pmatrix},
\end{equation}
where "$\times$" indicates non-zero component.
In this case we cannot obtain realistic quark mixing at renormalizable level.
We thus assume contributions from higher order operators
\begin{align}
\Delta L = & \frac{\delta Y_{a 3}^{u 1} }{\Lambda} \bar Q_{L_a} \tilde H_1 u_{R_3} \phi + \frac{\delta Y_{a 3}^{u 2} }{\Lambda} \bar Q_{L_a} \tilde H_2 u_{R_3} \phi
+ \frac{\delta Y_{3 a}^{u1} }{\Lambda} \bar Q_{L_3} \tilde H_1 u_{R_a} \phi^* + \frac{\delta Y_{3 a}^{u 2} }{\Lambda} \bar Q_{L_3} \tilde H_2 u_{R_a} \phi^*  \nonumber \\
& + \frac{\delta Y_{a 3}^{d 1} }{\Lambda} \bar Q_{L_a}  H_1 d_{R_3} \phi + \frac{\delta Y_{a 3}^{d 2} }{\Lambda} \bar Q_{L_a}  H_2 d_{R_3} \phi
+ \frac{\delta Y_{3 a}^{d 1} }{\Lambda} \bar Q_{L_3}  H_1 d_{R_a} \phi^* + \frac{\delta Y_{3 a}^{d 2} }{\Lambda} \bar Q_{L_3}  H_2 d_{R_a} \phi^*,
\label{eff_Yukawa}
\end{align}
where $a=1,2$ and we assume $U(1)_X$ charge of $\phi$ to be $Q_\phi = 3 Q_{12} - Q_H$ to make these terms gauge invariant.
We will discuss possible hidden sector realizing such non-renormalizable term in sec.~VI.

After electroweak symmetry breaking, the mass terms of the quark sector are 
\begin{equation}
-\mathcal{L}^{q}_{\rm M} = \Big( Y_1^u \frac{v}{\sqrt{2}}\cos\beta + Y_2^u \frac{v}{\sqrt{2}}\sin\beta \Big)_{ij}\bar{u}_{iL}u_{jR}
+ \Big( Y_1^d \frac{v}{\sqrt{2}}\cos\beta + Y_2^d \frac{v}{\sqrt{2}}\sin\beta \Big)_{ij}\bar{d}_{iL}d_{jR} + h.c.
\end{equation}
Then quark mass matrices are given by
\begin{eqnarray}
m^{u,d}_{ij} = (Y_1^{u,d})_{ij} \frac{v}{\sqrt{2}}\cos\beta + (Y_2^{u,d})_{ij} \frac{v}{\sqrt{2}}\sin\beta .
\end{eqnarray}
The mass matrices can be diagonalized by unitary matrices connecting flavour and mass eigenstates as $u^\prime_{L,R} = V^u_{L,R} u_{L,R}$ and $d^\prime_{L,R} = V^d_{L,R} d_{L,R}$
where we write primed field as mass eigenstates. 

Here we write Yukawa interactions in terms of mass eigenstates and the quark Yukawa interactions associated with $H^0,h^0,\xi^0$ and $A^0$ are 
\begin{eqnarray}
-\mathcal{L}_{u\bar{u}}^{H^0} &=& \Big[R_{11}H^0-\frac{R_{21}H^0}{\tan{\beta}}\Big]\bar{u}^\prime_LX^u u^\prime_R + \frac{R_{21}H^0}{v\sin{\beta}}\bar{u}^\prime_L m^{u^\prime}u^\prime_R + h.c \nonumber \\
-\mathcal{L}_{u\bar{u}}^{h^0} &=& \Big[R_{12}h^0 - \frac{R_{22}h^0}{\tan{\beta}}\Big]\bar{u}^\prime_L X^u u^\prime_R + \frac{R_{22}h^0}{v\sin{\beta}}\bar{u}^\prime m^{u^\prime}u^\prime_R + h.c \nonumber \\
-\mathcal{L}_{u\bar{u}}^{\xi^0} &=& \Big[R_{13}\xi^0-\frac{R_{23}\xi^0}{\tan{\beta}}\Big]\bar{u}^\prime_L X^u u^\prime_R + \frac{R_{23}\xi^0}{v\sin{\beta}}\bar{u}^\prime_L m^{u^\prime}u^\prime_R + h.c \nonumber \\
-\mathcal{L}^{A^{0}}_{u\bar{u}} &=& \Big[\frac{iA^0}{\sin\beta}\Big]\bar{u}^\prime_L X^u u^\prime_R - \frac{iA^0}{v\tan{\beta}}\bar{u}^\prime_L m^{u^\prime}u^\prime_R + h.c,
\end{eqnarray}
for up-type quarks and 
\begin{eqnarray}
-\mathcal{L}_{d\bar{d}}^{H^0} &=& \Big[-R_{11}H^0\tan{\beta} + R_{21}H^0\Big]\bar{d}^\prime_L X^d d^\prime_R + \frac{R_{11}H^0}{v\cos{\beta}}\bar{d}^\prime_L m^{d^\prime}d^\prime_R + h.c \nonumber \\
-\mathcal{L}_{d\bar{d}}^{h^0} &=& \Big[-R_{12}h^0\tan{\beta} + R_{22}h^0\Big]\bar{d}^\prime_L X^d d^\prime_R + \frac{R_{12}h^0}{v\cos{\beta}}\bar{d}^\prime_L m^{d^\prime}d^\prime_R + h.c \nonumber \\
-\mathcal{L}_{d\bar{d}}^{\xi^0} &=& \Big[-R_{13}\xi^0\tan{\beta} + R_{23}\xi^0\Big]\bar{d}^\prime_L X^d d^\prime_R + \frac{R_{13}\xi^0}{v\cos{\beta}}\bar{d}^\prime_L m^{d^\prime}d^\prime_R + h.c \nonumber \\
-\mathcal{L}^{A^{0}}_{d\bar{d}} &=& \Big[\frac{iA^0}{\cos\beta}\Big]\bar{d}'_LX^dd'_R - \frac{i\tan\beta A^0}{v}\bar{d}'_Lm^{d'}d'_R+h.c,
\end{eqnarray}
for down-type quarks.
Here the interaction matrix $X^{u,d}$ are defined by
\begin{equation}
X^u=V_L^u\frac{Y^u_1}{\sqrt{2}}V_R^{u^\dagger}, \quad
X^d=V_L^d\frac{Y^d_2}{\sqrt{2}}V_R^{d^\dagger}.
\end{equation}

\subsection{Lepton masses and Yukawa interactions}

Here we consider Yukawa couplings for leptons
\begin{eqnarray}
-\mathcal{L}^{yukawa}_{lepton}=\bar{L} Y^l_1 H_1 l_R + \bar{L} Y^l_2 H_2 l_R + h.c.
\end{eqnarray}
Under our $U(1)_X$ charge assignment leptons have flavour-universal charges and 
all the elements in $Y^l_{1,2}$ can be non-zero at renormalizable level.
After electroweak symmetry breaking we obtain charged lepton mass term such as
\begin{eqnarray}
-\mathcal{L}^{yukawa}_{lepton,mass} &=&
\Big[\frac{Y_1^l v\cos\beta}{\sqrt{2}} + \frac{Y_2^l v\sin\beta}{\sqrt{2}}\Big]_{ij}\bar{l}_{iL}l_{jR} + h.c. \nonumber \\
&=& m^l_{ij}\bar{l}_{iL}l_{jR} +h.c.
\end{eqnarray} 
The mass matrix can be diagonalized by biunitary transformation by defining the physical fields (mass eigenstates) of leptons as
$l^\prime_{L,R} = V^l_{L,R} l_{L,R}$, and diagonal mass matrix is given by $m^{l^\prime} = V^l_Lm^l V_R^{l\dagger}$.

The Yukawa interactions among neutral scalar and charged leptons can be written by
\begin{eqnarray}
-\mathcal{L}^{H^0}_{l\bar{l}} &=& \Big[R_{21}H^0-R_{11}H^0\tan{\beta}\Big]\bar{l}^\prime_L X^l l^\prime_R + \frac{R_{11}H^0}{v\cos{\beta}}\bar{l}^\prime_L m^{l^\prime}l^\prime_R + h.c \nonumber \\
-\mathcal{L}^{h^0}_{l\bar{l}} &=& \Big[R_{22}h^0 - R_{12}h^0\tan{\beta}\Big]\bar{l}^\prime_L X^l l^\prime_R + \frac{R_{12}h^0}{v\cos{\beta}}\bar{l}^\prime_L m^{l^\prime}l^\prime_R + h.c \nonumber \\
-\mathcal{L}^{\xi^0}_{l\bar{l}} &=& \Big[R_{23}\xi^0 - R_{13}\xi^0\tan{\beta}\Big]\bar{l}^\prime_L X^l l^\prime_R + \frac{R_{13}\xi^0}{v\cos{\beta}}\bar{l}^\prime_L m^{l^\prime}l^\prime_R + h.c \nonumber \\
-\mathcal{L}^{A^0}_{l\bar{l}} &=&  \Big[\frac{iA^0}{\cos\beta}\Big]\bar{l}^\prime_L X^l l^\prime_R  -\frac{i\tan\beta A^0}{v}\bar{l}^\prime m^{l^\prime}l^\prime_R + h.c.
\label{leptonic interaction to neutral scalars}
\end{eqnarray}
where coupling matrix $X^l$ is defined as 
\begin{eqnarray}
X^l=V_L^l\frac{Y^l_2}{\sqrt{2}}V_R^{l^\dagger}.
\end{eqnarray}
Note that $X^l$ is not diagonal in general which induce flavour violating processes which will be discussed below.

Finally, we consider the leptonic interaction with charged Higges, 
\begin{eqnarray}
-\mathcal{L}^{yukawa,H^\pm}_{lepton} = (Y_1^l)_{ij} \bar{\nu}_{iL}\phi^+_1 l_{jR} + (Y_2^l)_{ij}\nu_{iL}\phi_2^+ l_{jR} + h.c.
\end{eqnarray}
where mass eigenstate of the charged Higgs is given by Eq.~\eqref{Eq:chargedMES}.
In this paper, we consider neutrino mass is generated introducing right-handed neutrinos $\nu_R$ which are gauge singlet.
Then we can apply type-I seesaw mechanism where detailed analysis is omitted.
The physical $\nu_L^\prime$ is related to the unphysical $\nu_L$ by
\begin{eqnarray}
\nu^\prime_L=V_L^\nu \nu_L
\end{eqnarray}
Defining the Pontecorvo-Maki-Nakagawa-Sakata(PMNS) matrix, $V_{PMNS}$ as
\begin{eqnarray}
V_{PMNS}=V_L^\nu V_L^{l\dagger}
\end{eqnarray}
The leptonic interaction with charged Higgs takes the form
\begin{eqnarray}
-\mathcal{L}^{yukawa,H^\pm}_{lepton} &=& \sqrt{2}\bar{\nu}^\prime_L V_{PMNS}\Big[-\frac{\tan\beta}{v}m^{l^\prime} + \frac{X^l}{\cos\beta}\Big]l^\prime_R H^+ + h.c. \nonumber \\
&=& \bar{\nu}^\prime_L V_{PMNS} Y^l_{H^\pm} l^\prime_R H^+ + h.c.
\label{leptonic interaction to charged scalar}
\end{eqnarray}
where,
\begin{eqnarray}
Y^l_{H^\pm}= \sqrt{2}\Big[-\frac{\tan\beta}{v}m^{l^\prime} + \frac{X^l}{\cos\beta}\Big]
\end{eqnarray}

Writing the Yukawa interaction terms of the leptons, to the scalar, we get
\begin{eqnarray}
-\mathcal{L}^{yukawa}_{lepton} &=& \bar{l}^\prime_L\epsilon_\phi Y^l_\phi l^\prime_R \phi + \bar{\nu}^\prime_L V_{PMNS} Y^l_{H^\pm} l^\prime_R H^+ + h.c.
\end{eqnarray}
where $\phi=h^0,H^0,\xi^0,A^0$ and $\epsilon_{h^0,H^0,\xi^0}=1,\epsilon_{A^0}=i$. Also $Y^l_\phi$ can be obtained from Eq.~\ref{leptonic interaction to neutral scalars}.

\section{$Z'$ interaction with SM fermions and Atomki anomaly}

In this section, we discuss $Z'$ interactions associated with the SM fermions and possibility to explain Atomki anomaly.  \\
{\it Quark Sector:} Kinetic terms for quarks are 
\begin{eqnarray}
\mathcal{L}_{\text{quark}}=\bar{Q}^0_{jL}i\gamma_\mu D^\mu Q^0_{jL} + \bar{u}^0_{jR}i\gamma_\mu D^\mu u^0_{jR} + \bar{d}^0_{jR}i\gamma_\mu D^\mu d^0_{jR} 
\end{eqnarray}
where the superscrript "$0$" indicates the fermions in flavour eigenstate. The covariant derivatives are given by 
\begin{eqnarray}
D_{\mu} Q_{L_j}^0 &=& \left( \partial_\mu + i\frac{1}{2} g\tau^a W^a_\mu + ig^\prime \frac{1}{6} B_\mu + ig^{\prime\prime} Q_{Q_i}B^\prime_\mu \right) Q_{L_j}^0 \nonumber \\
D_{\mu} u_{R_j}^0 [d_{R_j}^0] &=& \left( \partial_\mu + ig^\prime \frac{2}{3} \left[ - \frac13 \right] B_\mu + ig^{\prime \prime} Q_{u_j[d_j]} B^\prime_\mu \right) u_{R_j}^0 [d_{R_j}^0] 
\end{eqnarray}
{\it Lepton Sector:} Kinetic terms for leptons are 
\begin{eqnarray}
\mathcal{L}_{\text{lepton}}=\bar{L}^0_{L_j}i\gamma_\mu D^\mu L^0_{L_j} + \bar{e}^0_{R_j}i\gamma_\mu D^\mu e^0_{R_j}  
\end{eqnarray}
where 
\begin{eqnarray}
D_{\mu} L^0_{L_j} &=& \left( \partial_\mu + i\frac{1}{2} g\tau^a W^a_\mu - ig^\prime \frac{1}{2} B_\mu + ig^{\prime\prime} Q_{L_j}B^\prime_\mu \right) L_{L_j} \nonumber \\
D_{\mu} &=& \left( \partial_\mu - ig^\prime B_\mu + ig^{\prime \prime} Q_{e_j} B^\prime_\mu \right) e_{R_j}
\end{eqnarray}

For discussing explanation of Atomki anomaly we focus on $Z'$ interactions associated with first generation of fermions.
The $Z^\prime_\mu$ coupling to first generation of quarks are given by
\begin{eqnarray}
\mathcal{L}\supset-\bar{u}\gamma^\mu(g_{V}^u -g_{A}^u \gamma_5)Z^\prime_\mu u + \bar{d}\gamma^\mu(g_{V}^d -g_{A}^d \gamma_5)Z^\prime_\mu d.
\end{eqnarray}
Applying Eq.~\eqref{Rw matrix} and \eqref{eq:gauge field}, the coupling coefficients are given by
\begin{eqnarray}
g_{V}^u &=& -\frac{1}{4} g \frac{\sin\theta}{\cos\theta_w} + \frac{2}{3} g \frac{\sin^2\theta_w}{\cos\theta_w}\sin\theta + \frac{5}{12}\rho g \frac{\sin\theta_w}{\cos\theta_w}\cos\theta - \frac{1}{2}g^{\prime\prime}\cos\theta\frac{\rho}{x}(Q_{Q_1} + Q_{u_1}) \nonumber \\
g_{A}^u &=& -\frac{1}{4}g\frac{\sin\theta}{\cos\theta_w} - \frac{1}{4}\rho g\frac{\sin\theta_w}{\cos\theta_w}\cos\theta - \frac{1}{2}g^{\prime\prime}\cos\theta\frac{\rho}{x}(Q_{Q_1}-Q_{u_1}) \nonumber \\
g_{V}^d &=& -\frac{1}{4}g\frac{\sin\theta}{\cos\theta_w} + \frac{1}{3}g\frac{\sin^2\theta_w}{\cos\theta_w}\sin\theta + \frac{1}{12}\rho g \frac{\sin\theta_w}{\cos\theta_w}\cos\theta + \frac{1}{2}g^{\prime\prime}\cos\theta \frac{\rho}{x}(Q_{Q_1} + Q_{d1}) \nonumber \\
g_{A}^d &=& -\frac{1}{4}g\frac{\sin\theta}{\cos\theta_w} - \frac{1}{4}\rho g \frac{\sin\theta_w}{\cos\theta_w}\cos\theta + \frac{1}{2}g^{\prime\prime}\cos\theta \frac{\rho}{x}(Q_{Q_1}-Q_{d1})
\label{Quark V-A couplings}
\end{eqnarray}

Similarly, the $Z^\prime_\mu$ coupling to first generation of leptons are given by
\begin{eqnarray}
\mathcal{L}\supset =-\bar{\nu}_e\gamma^\mu(g_V^{\nu_e} -g_A^{\nu_e}\gamma_5)Z^{\prime}_\mu \nu_e + \bar{e}\gamma^\mu(g^e_V - g_A^e \gamma_5)Z^\prime_\mu e
\end{eqnarray}
where the relevant coefficients are 
\begin{eqnarray}
g_{V}^{\nu_e} &=& g_{A}^{\nu_e} = -\frac{1}{4}g\frac{\sin\theta}{\cos\theta_w} - \frac{1}{4}\rho g \frac{\sin\theta_w}{\cos\theta_w}\cos\theta -\frac{1}{2}g^{\prime\prime}\cos\theta \frac{\rho}{x}Q_L \nonumber \\
g_{V}^e &=& -\frac{1}{4}g\frac{\sin\theta}{\cos\theta_w} + g\frac{\sin^2\theta_w}{\cos\theta_w}\sin\theta + \frac{3}{4}\rho g \frac{\sin\theta_w}{\cos\theta_w}\cos\theta + \frac{1}{2}g^{\prime\prime}\cos\theta \frac{\rho}{x}(Q_{L_1} + Q_{e_1}) \nonumber \\
g_{A}^e &=& -\frac{1}{4}g\frac{\sin\theta}{\cos\theta_w} - \frac{1}{4}\rho g \frac{\sin\theta_w}{\cos\theta_w}\cos\theta + \frac{1}{2}g^{\prime\prime}\cos\theta\frac{\rho}{x}(Q_{L_1}-Q_{e_1})
\label{Lepton V-A couplings}
\end{eqnarray}
The interaction between SM fermions and $Z^\prime_\mu$ can be written as 
\begin{align}
& \mathcal{L}^{Z^\prime} = -J^\mu_{Z^\prime} Z^\prime_\mu \\
& J^\mu_{Z^\prime} = e \sum_f \bar{\psi}_f(C^f_V + C^f_A \gamma_5)\psi_f
\end{align}
where $f = \{u, d, e, \nu_e \}$ distinguishes type of fermion.
Thus we have the relations
\begin{eqnarray}
C^{u, \nu_e}_V &=& \frac{g_V^{u, \nu_e}}{g \sin\theta_w}, \hspace{5mm} C^{u, \nu_e}_A = -\frac{g_A^{u, \nu_e}}{g \sin\theta_w} \nonumber \\
C^{d,e}_V &=& -\frac{g_V^{d,e}}{g \sin\theta_w}, \hspace{5mm} C^{d,e}_A = \frac{g_A^{d,e}}{g \sin\theta_w} 
\end{eqnarray}
To explain Atomki anomaly we require the constraints~\cite{Feng:2016jff,Feng:2016ysn}; 
\begin{eqnarray}
\mid C_{V}^p\mid &\lesssim& 1.2\times 10^{-3} \nonumber \\
\mid C_{V}^n\mid &=& (2-10)\times 10^{-3} \nonumber \\
\mid C_{V}^e\mid &=& (0.2-1.4)\times 10^{-3} \nonumber \\
\sqrt{C_{V}^e C_{V}^{\nu_e}} &\lesssim& 7\times 10^{-5}
\label{Atomki Constraints}
\end{eqnarray}
where the nucleon coupling to $Z^\prime_{\mu}$ are $C_{V}^p = 2 C_{V}^u + C_{V}^d$ and $C_{V}^n =  C_{V}^u + 2 C_{V}^d$.  
The $Z^\prime_{\mu}$ coupling needs to be protophobic and satisfies the relation
\begin{eqnarray}
-0.067 < \frac{C_{V}^p}{C_{V}^n}< 0.078
\label{Atomki Constraints 2}
\end{eqnarray}
The last constraint of Eq.~\eqref{Atomki Constraints} needs the vector coupling of electron neutrino to be very small. 
Now since for the case of neutrino the vector and axial couplings are equal, the axial coupling of neutrino can be made zero by the condition
\begin{eqnarray}
\frac{1}{2} g^{\prime\prime}\frac{\rho}{x}\cos\theta Q_H = \frac{1}{4}g \frac{\sin\theta}{\cos\theta_w} + \frac{1}{4}\rho g \frac{\sin\theta_w}{\cos\theta_w}\cos\theta
\label{Zero Axial coupling condition}
\end{eqnarray}
Here we assume the condition is at least satisfied approximately by tuning our free parameters.
Furthermore, the charge assignment conditions in Eq.~\eqref{charge equations 2} and Eq.~\eqref{Zero Axial coupling condition} makes all the axial couplings of $Z^{\prime}_{\mu}$ to be zero. 
Using the charges in Eq.~\eqref{charge equations 2}, we get 
\begin{eqnarray}
C_{V}^e = \sin\theta \frac{\cos\theta_w}{\sin\theta_w}
\end{eqnarray}
We define the parameters, $C_V^p=\delta$, $C_V^n=\epsilon$ and $C_{V}^e=\kappa$. Then 
\begin{eqnarray}
\kappa &=& \sin\theta \frac{\cos\theta_w}{\sin\theta_w} \nonumber \\
\delta &=& -\frac{1}{2}\frac{\sin\theta}{\sin\theta_w \cos\theta_w} + \frac{\sin\theta_w}{\cos\theta_w}\sin\theta + \frac{1}{2}\rho \frac{\cos\theta}{\cos\theta_w} -3\tilde{g}^{\prime\prime} \cos\theta\frac{\rho}{x}Q_{Q_{12}} \nonumber \\
\epsilon &=& \frac{1}{2}\frac{\sin\theta}{\sin\theta_w \cos\theta_w} + \frac{1}{2}\rho \frac{\cos\theta}{\cos\theta_w} - 3 \tilde{g}^{\prime\prime}\cos\theta\frac{\rho}{x}Q_{Q_{12}}
\end{eqnarray}
where $\tilde{g}^{\prime\prime}= g^{\prime\prime}/e = g^{\prime\prime}/g \sin\theta_w$. In terms of $\tilde{g}^{\prime\prime}$, Eq.~\eqref{Zero Axial coupling condition} can be written as 
\begin{eqnarray}
\frac{1}{2}\tilde{g}^{\prime\prime}\frac{\rho}{x}\cos\theta Q_H = \frac{1}{4}\frac{\sin\theta}{\sin\theta_w \cos\theta_w} + \frac{1}{4}\rho \frac{\cos\theta}{\cos\theta_w}
\label{Zero Axial coupling condition2}
\end{eqnarray}
The parameters $\kappa$, $\delta$ and $\epsilon$ are related as
\begin{eqnarray}
\delta = \epsilon - \kappa
\end{eqnarray}

 \begin{figure}[t!]
\includegraphics[width=80mm]{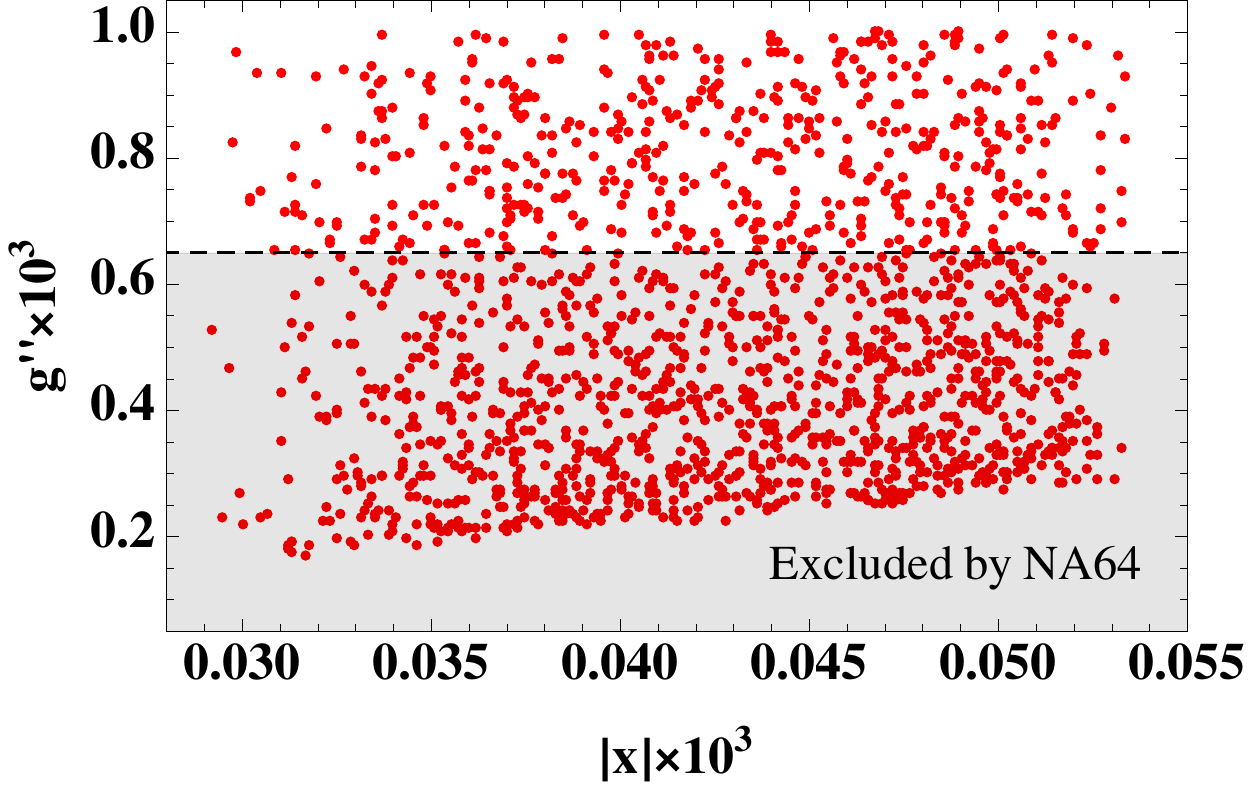}
\caption{Correlation between $|x|$ and $g''$ for the parameters satisfying the Atomki constraints.}
\label{fig:Atomki1}
\end{figure}

 \begin{figure}[t!]
\includegraphics[width=80mm]{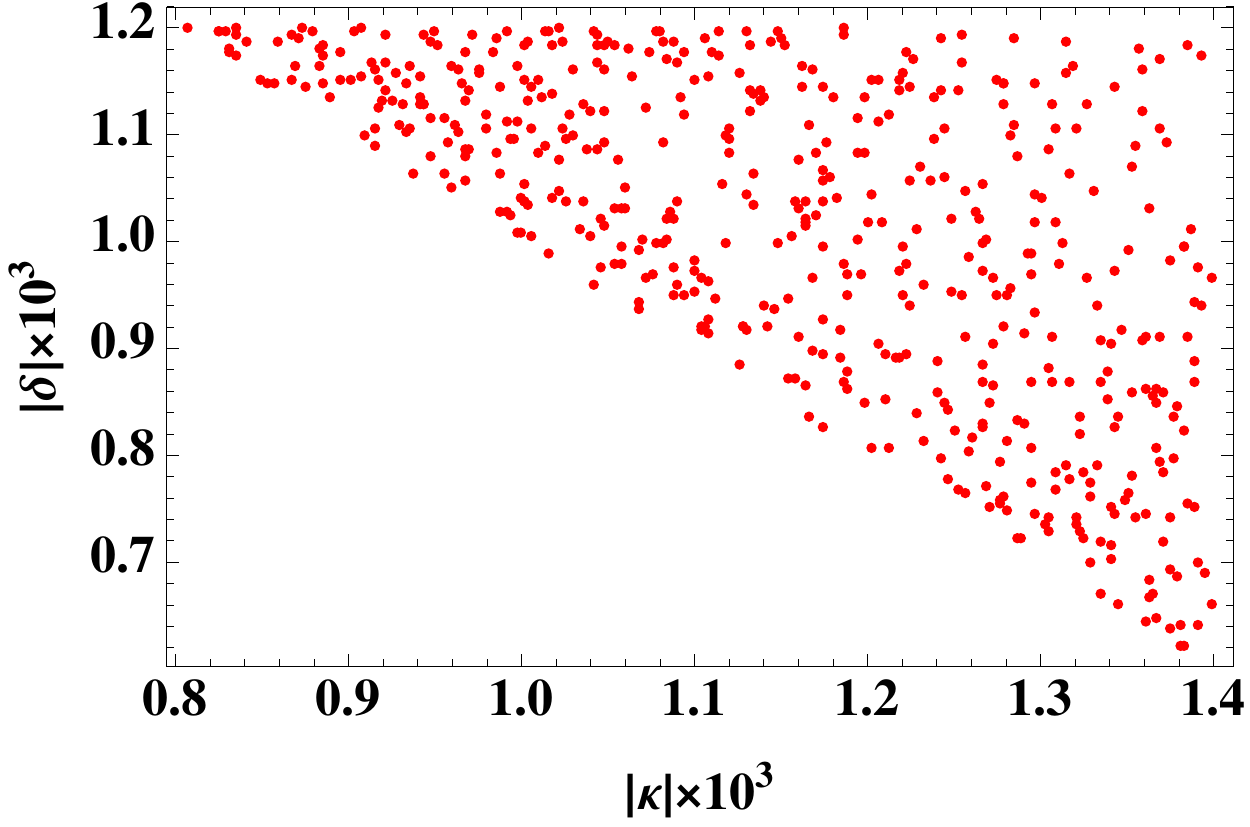}
\includegraphics[width=80mm]{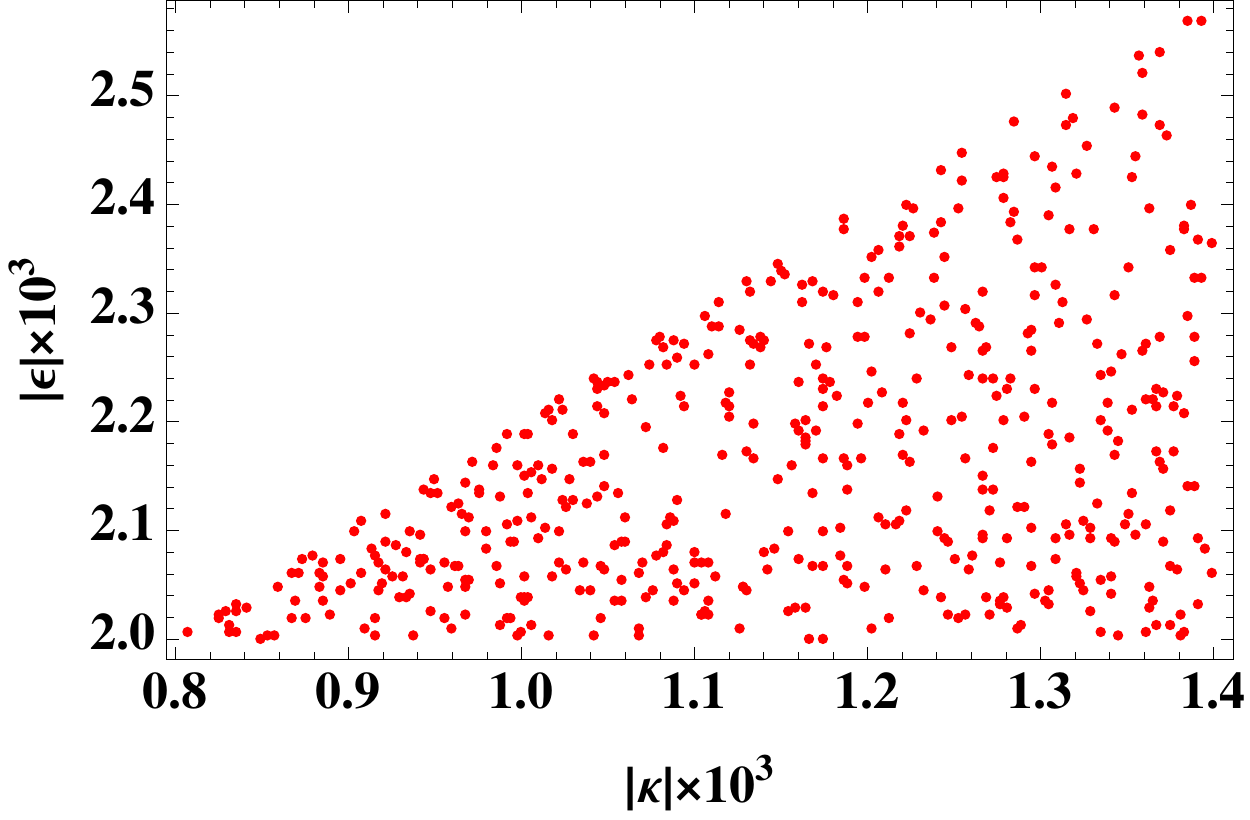} 
\caption{Left: correlation between $\kappa$ and $\delta$ for parameters satisfying the Atomki constraints. Right:  that of $\kappa$ and $\epsilon$. }
\label{fig:Atomki2}
\end{figure}

Here we comment on the case of flavour independent charge assignment $Q_{Q_{12}}=Q_{H}/3$ as given in Eq.~\eqref{charge equations 3}. 
In this case, the vector and axial couplings in Eq.~\eqref{Quark V-A couplings} become 
\begin{equation}
g_{V}^u = g\frac{\sin\theta}{\cos\theta_w}\Big(\frac{1}{3} \sin^2\theta_w - \frac{2}{3} \Big), \quad g_{V}^d = g\frac{\sin\theta}{\cos\theta_w}\Big( \frac{1}{3}\sin^2\theta_w -\frac{1}{3} \Big).
\end{equation}
Then it is not possible to satisfy Eq.~\eqref{Atomki Constraints}. 

We then scan parameters such that~\footnote{In fact only ratio of $Q_H$ and $Q_{12}$ is relevant free parameter where we can absorb one of them in normalizing $g''$. Here we simply scan both $Q_H$ and $Q_{12}$.}
\begin{align}
g'' \in [10^{-6}, 10^{-3}], \quad |x| \in [10^{-6}, 10^{-3}], \quad |Q_{H, 12}| \in [0.1, 1]
\end{align}
where kinetic mixing parameter $x$ and charges $Q_{H, 12}$ can be both positive and negative while $g''$ is taken to be positive.
We then impose conditions in Eq.~\eqref{Atomki Constraints} to explain Atomki anomaly. 
From the last condition we require 
\begin{equation}
\frac{1}{g \sin \theta_W} \left(\frac{1}{2} g^{\prime\prime}\frac{\rho}{x}\cos\theta Q_H - \frac{1}{4}g \frac{\sin\theta}{\cos\theta_w} - \frac{1}{4}\rho g \frac{\sin\theta_w}{\cos\theta_w}\cos\theta \right) <  3.5 \times 10^{-6},
\end{equation}
where the right hand side of the inequality is obtained by $7 \times 10^{-5}/$Max$[|C_V^e|] $.
Then the conditions Eq.~\eqref{Zero Axial coupling condition} and \eqref{Zero Axial coupling condition2} are approximately satisfied.
In Fig.~\ref{fig:Atomki1}, we show $g''$ and $\sin \theta$ values accommodating Atomki anomaly where shaded region is excluded by NA64 experiment~\cite{Banerjee:2019hmi} .  
We also show, in Fig.~\ref{fig:Atomki2}, resultant $|\kappa|$, $|\delta|$ and $|\epsilon|$ for parameters satisfying the Atomki constraints and NA64 limit for $g''$.
We find that $\sin \theta < 10^{-3}$ and it can be allowed by electroweak precision measurement~\cite{Langacker:2008yv}. 
Since typical value of $g''$ is $\mathcal{O}(10^{-3})$ the typical VEV of $\phi$ is estimated as $m_{Z'}(\simeq 17 {\rm MeV})/g'' \sim \mathcal{O}(10)$ GeV.
In the following scalar potential analysis we apply $v_\phi = 10$ GeV as a reference value.

Note that we have quark flavour changing interactions associated with $Z'$ since $U(1)_X$ charge assignment for quarks is flavour dependent 
where the interactions are suppressed by CKM elements~\cite{Pulice:2019xel}.
In addition such interactions are also induced by Yukawa interactions of two Higgs doublets and dark sector interactions shown in sec.VI at one-loop level.
Such interactions are constrained from $B^0$--$\bar B^0$ mixing~\cite{Becirevic:2016zri, Kumar:2019qbv} and meson decay $M \to M' Z'$ ($M$ and $M'$ indicate a meson) where 
the latter process would be enhanced due to light mass of $Z'$~\cite{Dror:2017nsg}.
In principle it is possible to avoid such constraints considering cancellation between contributions from tree level CKM suppressed $Z'$ interactions 
and those from two Higgs doublet and/or dark sector effect by tuning model parameters.
Detailed analysis of flavour changing interactions is beyond the scope of our work and we assume such interactions are suppressed.

\section{Analysis of scalar potential}

In this section, we analyze scalar potential of the model which includes two Higgs doublet and new scalar to break $U(1)_X$ gauge symmetry.
We write parameters in scalar potential 
$\{\lambda_1,\lambda_2,\lambda_3, \lambda_4, \lambda_{11},\lambda_{8},\lambda_{9} \}$
by physical masses and VEVs such that 
\begin{align}
\lambda_1 &= \frac{ m_{H^0}^2 R_{11}^2 + m_{h^0}^2 R_{12}^2 + m_{\xi^0}^2 R_{13}^2  - m_{H_{12}}^2 \frac{s_\beta}{c_\beta} }{2v^2 c_\beta^2} \\
\lambda_2 &= \frac{ m_{H^0}^2 R_{21}^2 + m_{h^0}^2 R_{22}^2 + m_{\xi^0}^2 R_{23}^2  - m_{H_{12}}^2 \frac{c_\beta}{s_\beta} }{2v^2 s_\beta^2} \\
\lambda_3 &= \frac{1}{v^2} \left( - \frac{m_{H_{12}}^2 }{c_\beta s_\beta} + 2 m_{H^\pm}^2 + \frac{ m_{H^0}^2 R_{11} R_{21} + m_{h^0}^2 R_{12} R_{22} + m_{\xi^0}^2 R_{13} R_{23} }{c_\beta s_\beta} \right) \\
\lambda_4 &= \frac{1}{v^2} \left( \frac{m_{H_{12}}^2 }{c_\beta s_\beta} + m_{A^0}^2 -2 m_{H^\pm}^2 \right) \\
\lambda_5 &= \frac{1}{v^2} \left( \frac{m_{H_{12}}^2 }{c_\beta s_\beta} - m_{A^0}^2  \right) \\
\lambda_{11} &= \frac{ m_{H^0}^2 R_{31}^2 + m_{h^0}^2 R_{32}^2 + m_{\xi^0}^2 R_{33}^2 }{2 v_\phi^2 } \\
\lambda_8 &= \frac{ m_{H^0}^2 R_{11} R_{31} + m_{h^0}^2 R_{12} R_{32} + m_{\xi^0}^2 R_{13} R_{33}   }{v v_\phi c_\beta } \\
\lambda_9 &= \frac{ m_{H^0}^2 R_{21} R_{31} + m_{h^0}^2 R_{22} R_{32} + m_{\xi^0}^2 R_{23} R_{33}   }{v v_\phi s_\beta }. 
\end{align}

The constraints from unitarity and perturbativity are given by~\cite{Bian:2017xzg,Muhlleitner:2016mzt} 
\begin{align}
& |\lambda_{1,2,3,11 }| \leq 4 \pi, \quad |\lambda_{8,9}| \leq 8 \pi, \quad |\lambda_{3} \pm \lambda_4| \leq 8 \pi, \quad  |\lambda_3 + 2 \lambda_4 + 3 \lambda_5| \leq 8 \pi, \nonumber \\
& \sqrt{|\lambda_3 (\lambda_3 + 2 \lambda_4)|} \leq 8 \pi,  \quad
\left| \lambda_1 + \lambda_2 \pm \sqrt{(\lambda_1 - \lambda_2)^2 + \lambda^2_4} \right| \leq 8 \pi, \nonumber \\
& \left| \lambda_1 + \lambda_2 \pm \sqrt{(\lambda_1 - \lambda_2)^2 + \lambda^2_5} \right| \leq 8 \pi, \quad a_{1,2,3} \leq 8 \pi,
\end{align}
where $a_{1,2,3}$ are the solution of the following equation
\begin{align}
& x^3 - 2x^2 (3 \lambda_1 + 3 \lambda_2 + 2 \lambda_{11})  \nonumber \\
& -  x (2 \lambda_{8}^2 + 2 \lambda_{9}^2 - 36 \lambda_1 \lambda_2 - 24 \lambda_1 \lambda_{11} - 24 \lambda_1 \lambda_{11} + 4 \lambda_3^2 + 4\lambda_3 \lambda_4 + \lambda_4^2 ) \nonumber \\
& + 4 (3 \lambda_{8}^2 \lambda_2 - \lambda_{8} \lambda_{9} (2 \lambda_3 + \lambda_4) + 3 \lambda_{9}^2 \lambda_1 + \lambda_{11} ((2 \lambda_3 + \lambda_4)^2 - 36 \lambda_1 \lambda_2)) =0.
\end{align}
We adopt the conditions for vacuum stability~\cite{Muhlleitner:2016mzt}:
\begin{align}
& \Omega_1 \cup \Omega_2 \\
& \Omega_1 = \biggl\{ \lambda_{1,2,11} >0, \ 2 \sqrt{\lambda_1 \lambda_{11} } + \lambda_{8} >0, \ 2 \sqrt{\lambda_2 \lambda_{11} } + \lambda_{9} >0,   \nonumber \\
& \qquad \qquad 2\sqrt{\lambda_1 \lambda_2} + \lambda_3 > 0, \  \lambda_8 + \sqrt{\frac{\lambda_1}{\lambda_2}} \lambda_8  \geq 0 \biggr\} \\
& \Omega_2 = \biggl\{ \lambda_{1,2,11} >0, \ 2 \sqrt{\lambda_2 \lambda_{11} } \geq \lambda_{9}  > -2 \sqrt{\lambda_2 \lambda_{11} }, 
\ 2 \sqrt{\lambda_1 \lambda_{11} } > -\lambda_8  \geq \sqrt{\frac{\lambda_1}{\lambda_2}} \lambda_9,   \nonumber \\
& \qquad \qquad \sqrt{(\lambda_8^2 - 4 \lambda_1 \lambda_{11})(\lambda_9^2 - 4 \lambda_2 \lambda_{11}) } > \lambda_8 \lambda_9 - 2 \lambda_3 \lambda_{11} \biggr\}
\label{vac}
\end{align}
where we assume $\lambda_4 \leq \lambda_5$ by choosing $m_{A^0} < m_{H^\pm}$.
In addition, we consider constraints from $h^0 \to Z' Z'$ decay where the decay width of the process is approximately given by
\begin{equation}
\Gamma_{h^0 \to Z' Z'} \simeq \frac{1}{32 \pi} \frac{m_{h^0}^3}{v_\phi^2} R_{32}^2.
\end{equation}
Since $Z'$ is very light as $m_{Z'} \sim 17$ MeV it is difficult to see decay products from the decay chain, and we consider the mode contributes to Higgs invisible decay mode.
In addition, we consider $h^0 \to \xi^0 \xi^0$ decay which also contribute to Higgs invisible decay mode since $\xi^0$ mainly decays into $Z'Z'$ pair.
The decay width is given by
\begin{eqnarray}
 \Gamma_{h^0 \to \xi^0 \xi^0} &=& \frac{C_{h^0 \xi^0 \xi^0}^2}{16 \pi m_{h^0}} \sqrt{1 - \frac{4 m_{\xi^0}^2}{m_{h^0}^2}}, \\
C_{h^0 \xi^0 \xi^0} &=& R_{12}(6 R_{13}^2 v_1 \lambda_1 + v_1 (R_{23}^2 \tilde{\lambda} + R_{33}^2 \lambda_8) + 2R_{13}(R_{23}v_2 \tilde{\lambda} +R_{33} v_{\phi}\lambda_8)) \nonumber \\
&+& R_{32}(6 R_{33}^2 v_{\phi}\lambda_{11} + 2 R_{13}R_{33}v_1 \lambda_8 + R_{13}^2 v_{\phi}\lambda_8 + 2 R_{23} R_{33}v_2 \lambda_9  \nonumber \\
 &+& R_{23}^2 v_{\phi}\lambda_9) + R_{22}(6 R_{23}^2 v_2 \lambda_2 + v_2(R_{13}^2 \tilde{\lambda} + R_{33}^2 \lambda_9) \nonumber \\
 &+& 2 R_{23}(R_{13}v_1 \tilde{\lambda} + R_{33}v_{\phi} \lambda_9)),
\end{eqnarray}
Then we apply constraint from invisible decay of the SM Higgs for the branching ratio of the process~\cite{Sirunyan:2018koj}
\begin{equation}
BR(h^0 \to Z' Z') + BR(h^0 \to \xi^0 \xi^0) < 0.23.
\end{equation}
 In addition to the invisible decay BR, the other SM Higgs signal measurements also constrain the mixings among scalar bosons 
since it induces deviation of the SM Higgs couplings~\cite{ATLAS:2018doi, Sirunyan:2018koj}.
In our analysis we apply {\it HiggsSignals-2.2.3beta}~\cite{Bechtle:2013xfa} code to impose the constraints from the Higgs signal strength data and to obtain the $2\sigma$ allowed parameter space.

Note also that we can consider flavour constraints such as $b \to s \gamma$ induced from quark Yukawa couplings. 
In this paper, we do not discuss such constraints and we assume quark Yukawa couplings are tuned to satisfy all the quark flavour constraints. 
For lepton Yukawa couplings, we discuss constrains from flavour violation along with muon $g-2$ in Sec.~\ref{sec:LFV}.

 \begin{figure}[t!]
\includegraphics[width=80mm]{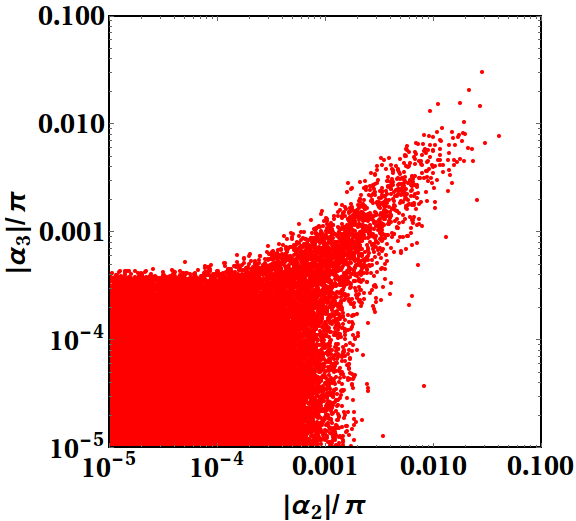}
\includegraphics[width=76mm]{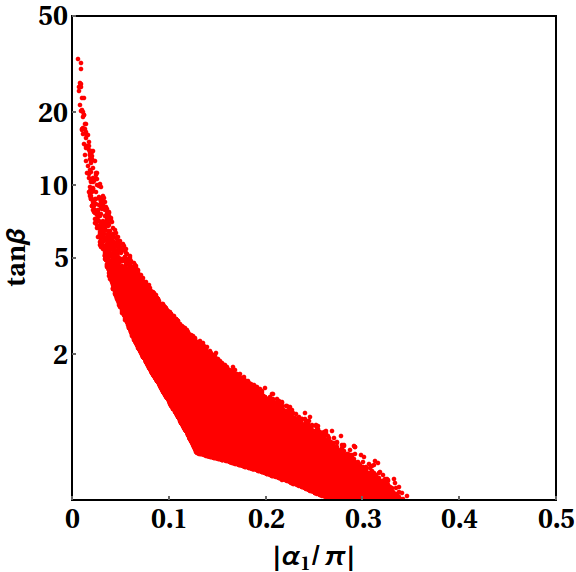} 
\includegraphics[width=78mm]{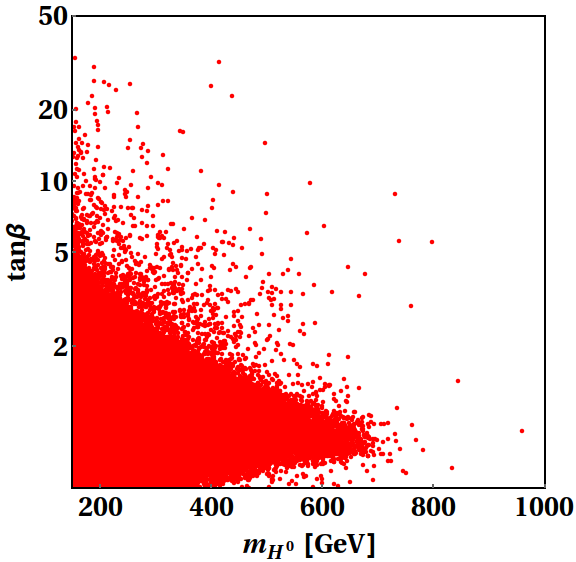}
\includegraphics[width=76mm]{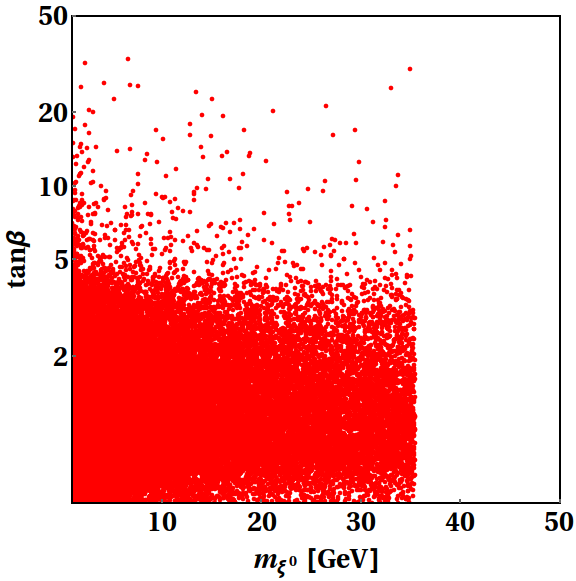}
\caption{The allowed parameter regions satisfying constraints from scalar potential and Higgs measurements.}
\label{fig:allowed1}
\end{figure}

 \begin{figure}[t!]
\includegraphics[width=80mm]{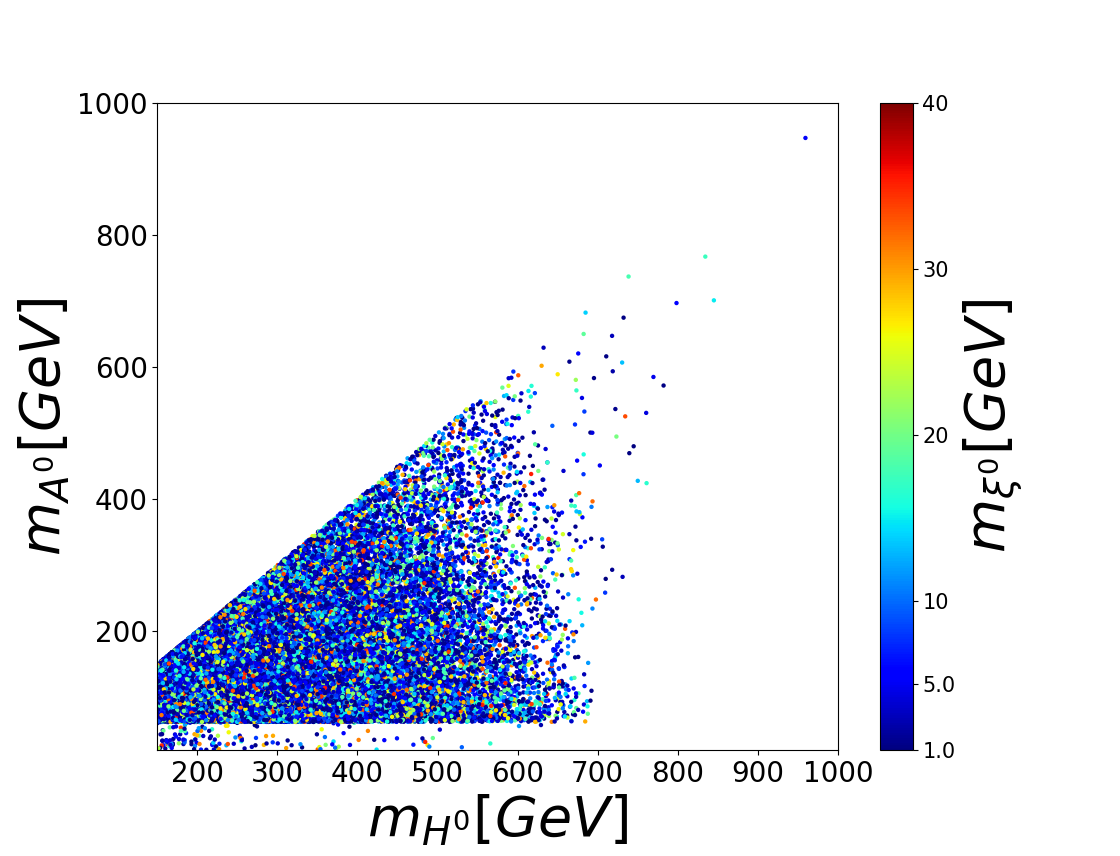}
\includegraphics[width=80mm]{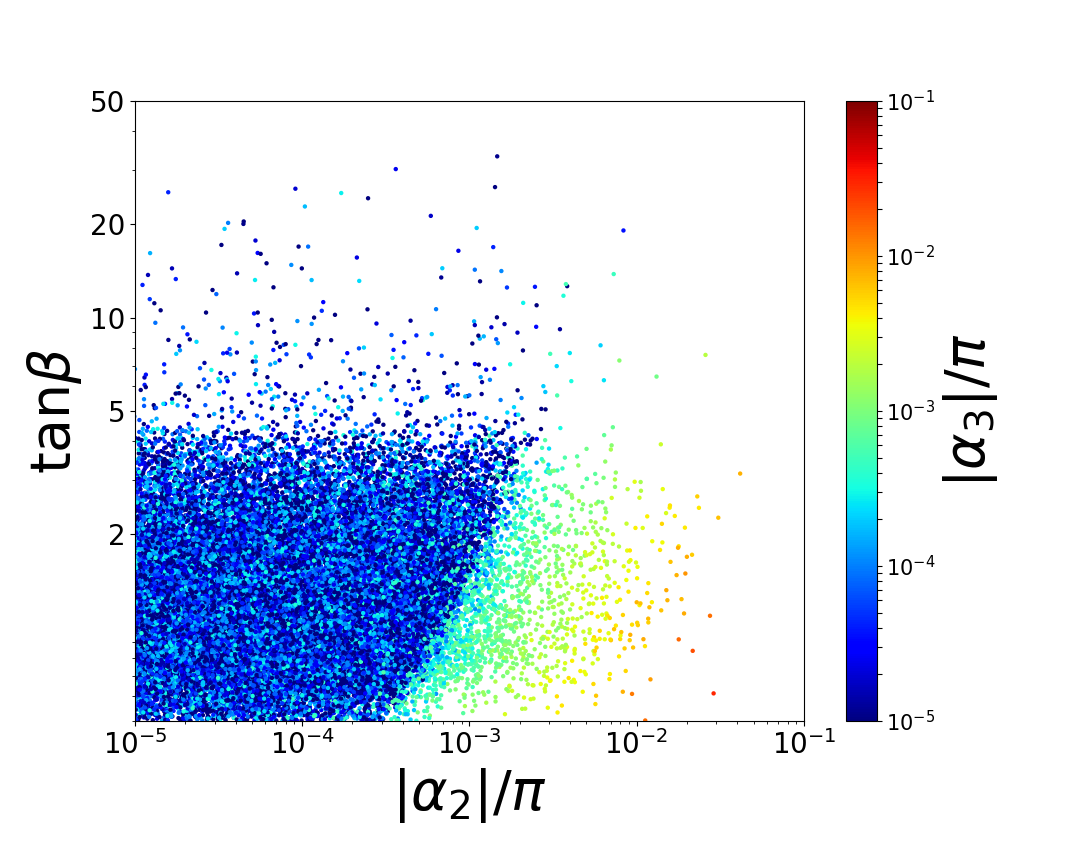}
\caption{The allowed scalar mass regions satisfying constraints from scalar potential and Higgs measurements.}
\label{fig:allowed2}
\end{figure}

Here we scan out parameters to search for allowed parameter region, such that
\begin{align}
& \alpha_{1,2,3} \in \left[-\frac{\pi}{2}, \frac{\pi}{2} \right], \quad m_{H^0} = m_{H^\pm} \in [150, 1000] \ {\rm [GeV]}, \quad m_{A^0} \in [20, 500] \ {\rm [GeV]}, \nonumber \\
& m_{\xi^0} \in [1, 500] \ {\rm [GeV]}, \quad \tan \beta \in [0.5, 50], \quad |m_{H_{12}}^2| \in [10, 10^6] \ {\rm [GeV^2]},
\label{parameter range scalar}
\end{align} 
$v_\phi = 10$ GeV is applied as indicated by solving Atomki anomalies,
and we assume $m_{A^0} < m_{H^0}$ for simplicity.
Further, $m_{H^0} = m_{H^\pm}$ is assumed to avoid constraints from the $T$-parameter.
We show the allowed parameter regions in Fig.~\ref{fig:allowed1} for mixing angles, $\tan \beta$ and scalar masses.
It is then found that $\alpha_1 > 0$ is preferred and there is correlation between $\tan \beta$ and $\alpha_1$.
Furthermore $\alpha_2$ and $\alpha_3$ should be small to suppress BR of $h^0 \to Z' Z'$ and $h^0 \to \xi^0 \xi^0$ processes.
The mass of $\xi^0$ is found to be less than $\sim 35$ GeV due to small VEV of $\phi$ and perturbativity constraints for couplings in scalar potential.
We also show correlations among scalar masses for allowed parameter region in Fig.~\ref{fig:allowed2}. For $m_{A^0}< m_{h^0}/2$, the parameter space gets strongly constrainted as the decay width of $h^0 \to A^0 A^0$ becomes significantly high.

Before closing this section, we comment on collider signals/constraints from additional Higgs bosons.
As in the THDMs extra Higgs bosons $H^\pm$, $A^0$ and $H^0(\xi^0)$ can be produced at the LHC and we expect signals of them. 
The decay BRs of them depend on Yukawa couplings and scalar mixing angles. 
It is quite complicated to list all the BRs since the Yukawa couplings are general one in our scenario and it is beyond the scope of this paper. 
Detailed discussions regarding extra Higgs boson signals can be referred to e.g. refs.~\cite{Sanyal:2019xcp,Chen:2018uim,Aggleton:2016tdd,Keus:2015hva,Arhrib:2015maa,Dumont:2014wha, Celis:2013ixa,Bernon:2014nxa,Craig:2015jba,Bernon:2015qea,Kling:2020hmi,Aiko:2020ksl}.
Note also that $H^0$ and $A^0$ can decay into $\xi^0 \xi^0$ mode since it tends to be light.
As $\xi^0$ dominantly decays into $Z'Z'$ we would have decay chain of $H^0(A^0) \to \xi^0 \xi^0 \to 4Z'$ where $Z'$ decays into the SM fermions.
Since $Z'$ is so light as $\sim 17$ MeV to explain Atomki anomaly it is very challenging to identify such $Z'$ at detectors of the LHC.
We thus do not discuss such signals although it will be the unique signatures of our scenario.

\section{Muon $g-2$ and Lepton Flavour Violation \label{sec:LFV}}

In this section, we discuss muon $g-2$ and constraints from LFV processes induced by Yukawa interactions in Eq.~\eqref{leptonic interaction to neutral scalars}.

In our model we obtain one loop diagrams contributing to muon $g-2$ in which neutral scalar boson and charged leptons propagating. 
Among them diagrams including $\tau$ inside loop would be dominant since they have enhanced by $m_\tau$ via chiral flip, 
and they are proportional to $X^\ell_{23} X^\ell_{32}$ factor.
Then the muon $g-2$ contribution from such diagrams is 
\begin{eqnarray}
\Delta a_\mu &=& \frac{m_\mu m_\tau X^l_{23}X^l_{32}}{8\pi^2}\sum_{\phi=h^0,H^0,\xi^0,A^0}Z_{\phi}
\label{muon g-2 1}
\end{eqnarray}
where explicit forms of $Z_{\phi}$ are given by 
\begin{align}
& Z_{h^0} \simeq (R_{22}-R_{12}\tan{\beta})^2\frac{\left[ \log{\frac{m_{h^0}^2}{m^2_{\tau}}}-\frac{3}{2}\right]}{m^2_{h^0}}, \quad 
Z_{H^0} \simeq (R_{21}-R_{11}\tan{\beta})^2\frac{\left[\log{\frac{m_{H^0}^2}{m^2_{\tau}}}-\frac{3}{2}\right]}{{m^2_{H^0}}}, \nonumber \\
& Z_{\xi^0} \simeq (R_{23}-R_{13}\tan{\beta})^2\frac{\left[\log{\frac{m_{\xi^0}^2}{m^2_{\tau}}}-\frac{3}{2}\right]}{{m^2_{\xi^0}}}, \quad
Z_{A^0} \simeq -\frac{1}{\cos^2\beta}\frac{\left[\log{\frac{m_{A^0}^2}{m^2_{\tau}}}-\frac{3}{2}\right]}{{m^2_{A^0}}}.
\label{muon g-2 2}
\end{align}
Here we neglect all the terms quadratic in muon mass as they are subdominant, hence the contribution from charged Higgs interactions is neglected. 
 In addition, we assume scalar boson masses are sufficiently heavier than charged lepton masses.

Since muon $g-2$ is explained by flavour violating coupling $X_{23,32}^\ell$ we need to take into account constraints from LFV processes.
Firstly we consider $h^0 \to \mu \tau$ decay, and the branching ratio is given by 
\begin{eqnarray}
BR(h^0\rightarrow\mu\tau)=(R_{22}-R_{12}\tan{\beta})^2(|X^l_{23}|^2+|X^l_{32}|^2)\frac{m_{h^0}}{16\pi\Gamma_{h^0}}
\label{Higgs to mu tau}
\end{eqnarray}
where $m_{h^0}=125$ GeV and we assume $\Gamma_{h^0}=4.21$ MeV. 

Secondly we consider $\tau \to 3 \mu$ process.
The branching ratio of the process is given by
\begin{eqnarray}
BR(\tau\rightarrow 3\mu) &=& \frac{\tau_\tau m_\tau^5}{3\times2^9\pi^3}\Big|X^l_{23}\Big|^2\Big[\Big|(R_{22}-R_{12}  \tan{\beta})
\frac{(Y_{h^0}^l)_{22}}{m_{h^0}^2} + \frac{(R_{21}-R_{11}\tan{\beta})(Y_{H^0}^l)_{22} }{m^2_{H^0}} \nonumber \\ 
&+& \frac{(R_{23}-R_{13}\tan{\beta})(Y_{\xi^0}^l)_{22}}{{m^2_{\xi^0}}}\Big|^2+\frac{(Y^l_{A^0})_{22}^2}{\cos^2\beta m^4_{A^0}}\Big],
\label{tau to 3mu}
\end{eqnarray}
where $\tau_\tau$ is the lifetime of $\tau$.

Thirdly we consider $\mu \to e \gamma$ process where the effective Lagrangian can be written as 
\begin{eqnarray}
\mathcal{L}_{\mu \rightarrow e\gamma} = \frac{e m_\mu}{16\pi^2}\bar{e}\sigma_{\mu\nu}(C_L P_L + C_R P_R)\mu F^{\mu\nu}.
\end{eqnarray}
The ratio of the muon branching ratios given by
\begin{eqnarray}
\frac{BR(\mu\rightarrow e\gamma)}{Br(\mu\rightarrow e\bar{\nu}_e \nu_\mu)}=\frac{3\alpha_e}{4\pi G^2_F}\big(\big|C_L\big|^2+\big|C_R\big|^2\big)
\label{mu to e gam}
\end{eqnarray}
where $G_F$ is the Fermi constant.
The coefficients are sum of contributions from scalar boson loop diagrams
\begin{eqnarray}
C_{L(R)}=C_{L(R)}^\phi+C_{L(R)}^{H^\pm}\\
C_L^{\phi}=\sum_{i=h^0,H^0,\xi^0,A^0}C_L^{\phi_i},
\end{eqnarray}
where the explicit forms are given by
\begin{eqnarray}
C_R^{h^0} &=& C_L^{h^0} \simeq \frac{X_{32}^lX_{13}^l}{2}\frac{m_\tau}{m_\mu}(R_{22}-R_{12}\tan{\beta})^2\frac{\Big[\log{\frac{m_{h^0}^2}{m^2_{\tau}}}-\frac{3}{2}\Big]}{m^2_{h^0}}, \nonumber \\
C_R^{H^0} &=& C_L^{H^0} \simeq \frac{X_{32}^lX_{13}^l}{2}\frac{m_\tau}{m_\mu}(R_{21}-R_{11}\tan{\beta})^2\frac{\Big[\log{\frac{m_{H^0}^2}{m^2_{\tau}}}-\frac{3}{2}\Big]}{m^2_{H^0}}, \nonumber \\
C_R^{\xi^0} &=& C_L^{\xi^0} \simeq \frac{X_{32}^lX_{13}^l}{2}\frac{m_\tau}{m_\mu}(R_{23}-R_{13}\tan{\beta})^2\frac{\Big[\log{\frac{m_{\xi^0}^2}{m^2_{\tau}}}-\frac{3}{2}\Big]}{m^2_{\xi^0}}, \nonumber \\
C_R^{A^0} &=& C_L^{A^0} \simeq -\frac{X_{32}^lX_{13}^l}{2}\frac{m_\tau}{m_\mu}\frac{1}{\cos^2\beta}\frac{\Big[\log{\frac{m_{A^0}^2}{m^2_{\tau}}}-\frac{3}{2}\Big]}{m^2_{A^0}},
\end{eqnarray}
and 
\begin{eqnarray}
C_L^{H^\pm} &\equiv& - \frac{1}{12 m^2_{H^{\pm}}} \Big(\frac{2X^l_{23}X^l_{13}}{\cos^2\beta}\Big) \nonumber \\
C_R^{H^\pm} &=& 0.
\end{eqnarray}
 Here we also show approximated forms assuming scalar bosons are sufficiently heavier than charged lepton masses.

Finally we consider $\tau \to \mu \gamma$ process.
The relevant effective Lagrangian is 
\begin{eqnarray}
\mathcal{L}_{\tau \rightarrow \mu\gamma} = \frac{e}{16\pi^2}m_{\tau}\bar{\mu}\sigma_{\mu\nu}(C^\prime_L P_L + C^\prime_R P_R)\tau F^{\mu\nu}
\end{eqnarray} 
where the Wilson coefficients are sum of contributions from the one-loop neutral and charged Higgs bosons
\begin{eqnarray}
C_{L(R)}^\prime = \sum_{\phi=h^0,H^0,\xi^0,A^0}C^{\prime \phi}_{L(R)} + C^{\prime H^\pm}_{L(R)}.
\end{eqnarray}
The explicit forms of the coefficients are given by
\begin{eqnarray}
C^{\prime h^0}_R &=& C^{\prime h^0}_L \simeq \frac{1}{2}\frac{(R_{22} - R_{12}\tan\beta)}{m_{h^0}^2}X^l_{32}(Y^l_{h^0})_{33}\Big[\log\frac{m^2_{h^0}}{m^2_\tau} - \frac{4}{3}\Big] \nonumber \\
C^{\prime H^0}_R &=& C^{\prime H^0}_L \simeq \frac{1}{2}\frac{(R_{21} - R_{11}\tan\beta)}{m_{H^0}^2}X^l_{32}(Y^l_{H^0})_{33}\Big[\log\frac{m^2_{H^0}}{m^2_\tau} - \frac{4}{3}\Big] \nonumber \\
C^{\prime \xi^0}_R &=& C^{\prime \xi^0}_L \simeq \frac{1}{2}\frac{(R_{23} - R_{13}\tan\beta)}{m_{\xi^0}^2}X^l_{32}(Y^l_{\xi^0})_{33}\Big[\log\frac{m^2_{\xi^0}}{m^2_\tau} - \frac{4}{3}\Big] \nonumber \\
C^{\prime A^0}_R &=& C^{\prime A^0}_L \simeq -\frac{1}{2}\frac{1}{\cos\beta m_{A^0}^2}X^l_{32}(Y^l_{A^0})_{33}\Big[\log\frac{m^2_{A^0}}{m^2_\tau} - \frac{5}{3}\Big] 
\end{eqnarray}
and 
\begin{eqnarray}
C^{\prime H^\pm}_L &\simeq& - \frac{1}{12 m^2_{H^\pm}}\Big(\frac{\sqrt{2}X^l_{32}}{\cos\beta}(Y^l_{H^\pm})_{33}\Big) \nonumber \\
C^{\prime H^\pm}_R &=& 0,
\end{eqnarray}
where we assume scalar boson masses are sufficiently heavier than charged lepton masses.
In addition we also include the two-loop contributions given by
\begin{eqnarray}
C^{h^0 t(b)}_{2R} &=& C^{h^0 t(b)}_{2L} = 2(R_{22} - R_{12}\tan\beta)X^l_{32}(Y_{h^0}^{u(d)})_{33} \frac{N_c Q_f^2 \alpha_e}{\pi}\frac{1}{m_\tau m_{t(b)}}f\Big(\frac{m^2_{t(b)}}{m^2_{h^0}}\Big) \nonumber \\
C^{H^0 t(b)}_{2R} &=& C^{H^0 t(b)}_{2L} = 2(R_{21} - R_{11}\tan\beta)X^l_{32}(Y_{H^0}^{u(d)})_{33} \frac{N_c Q_f^2 \alpha_e}{\pi}\frac{1}{m_\tau m_{t(b)}}f\Big(\frac{m^2_{t(b)}}{m^2_{H^0}}\Big) \nonumber \\
C^{\xi^0 t(b)}_{2R} &=& C^{\xi^0 t(b)}_{2L} = 2(R_{23} - R_{13}\tan\beta)X^l_{32}(Y_{\xi^0}^{u(d)})_{33} \frac{N_c Q_f^2 \alpha_e}{\pi}\frac{1}{m_\tau m_{t(b)}}f\Big(\frac{m^2_{t(b)}}{m^2_{\xi^0}}\Big) \nonumber \\
C^{A^0 t(b)}_{2R} &=& C^{A^0 t(b)}_{2L} = -2 \frac{X^l_{32}(Y_A^{u(d)})_{33}}{\cos\beta}\frac{N_c Q_f^2 \alpha_e}{\pi}\frac{1}{m_\tau m_{t(b)}}f\Big(\frac{m^2_{t(b)}}{m^2_{A^0}}\Big) 
\end{eqnarray}
\begin{eqnarray}
C^{W,h^0}_{2R} &=& C^{W,h^0}_{2L} = -(R_{22}-R_{12}\tan\beta)X^l_{32}(R_{12}\cos\beta + R_{22}\sin\beta)\frac{g\alpha_e}{2\pi m_\tau m_W}\nonumber \\
&& \Big[3f\Big(\frac{m_W^2}{m^2_{h^0}}\Big) + \frac{23}{4} g\Big(\frac{m_W^2}{m^2_{h^0}}\Big) + \frac{3}{4}h\Big(\frac{m_W^2}{m^2_{h^0}}\Big) + \frac{m^2_{h^0}}{2 m_W^2}\Big\lbrace f\Big(\frac{m_W^2}{m^2_{h^0}}\Big) - g\Big(\frac{m_W^2}{m^2_{h^0}}\Big)  \Big\rbrace\Big] \nonumber \\
C^{W,H^0}_{2R} &=& C^{W,H^0}_{2L} = -(R_{21}-R_{11}\tan\beta)X^l_{32}(R_{11}\cos\beta + R_{21}\sin\beta)\frac{g\alpha_e}{2\pi m_\tau m_W}\nonumber \\
&& \Big[3f\Big(\frac{m_W^2}{m^2_{H^0}}\Big) + \frac{23}{4} g\Big(\frac{m_W^2}{m^2_{H^0}}\Big) + \frac{3}{4}h\Big(\frac{m_W^2}{m^2_{H^0}}\Big) + \frac{m^2_{H^0}}{2 m_W^2}\Big\lbrace f\Big(\frac{m_W^2}{m^2_{H^0}}\Big) - g\Big(\frac{m_W^2}{m^2_{H^0}}\Big)  \Big\rbrace\Big] \nonumber \\
C^{W,\xi^0}_{2R} &=& C^{W,\xi^0}_{2L} = -(R_{23} - R_{13}\tan\beta)X^l_{32}(R_{13}\cos\beta + R_{23}\sin\beta)\frac{g\alpha_e}{2\pi m_\tau m_W}\nonumber \\
&& \Big[3f\Big(\frac{m_W^2}{m^2_{\xi^0}}\Big) + \frac{23}{4} g\Big(\frac{m_W^2}{m^2_{\xi^0}}\Big) + \frac{3}{4}h\Big(\frac{m_W^2}{m^2_{\xi^0}}\Big) + \frac{m^2_{\xi^0}}{2 m_W^2}\Big\lbrace f\Big(\frac{m_W^2}{m^2_{\xi^0}}\Big) - g\Big(\frac{m_W^2}{m^2_{\xi^0}}\Big) \Big\rbrace\Big] \nonumber \\
\end{eqnarray}
The loop functions are given as 
\begin{eqnarray}
f(z) &=& \frac{z}{2} \int_0^1 dx \frac{(1-2x(1-x))}{x(1-x)-z}\log\frac{x(1-x)}{z} \nonumber \\
g(z) &=& \frac{z}{2} \int_0^1 dx \frac{1}{x(1-x)-z}\log\frac{x(1-x)}{z} \nonumber \\
h(z) &=& - \frac{z}{2} \int_0^1 dx \frac{1}{x(1-x)-z}\Big[ 1 - \frac{z}{x(1-x)-z} \log \frac{x(1-x)}{z} \Big]
\end{eqnarray}
Then the branching ratio for $\tau \rightarrow \mu \gamma$ is expressed as
\begin{eqnarray}
\frac{BR(\tau \rightarrow \mu \gamma)}{BR(\tau \rightarrow \mu\bar{\nu}_\mu \tau)} = \frac{3\alpha_e}{4\pi G_F^2}(\mid C^\prime_L \mid^2 + \mid C^\prime_R \mid^2)
\end{eqnarray}


\begin{figure}[t!]
\includegraphics[width=80mm]{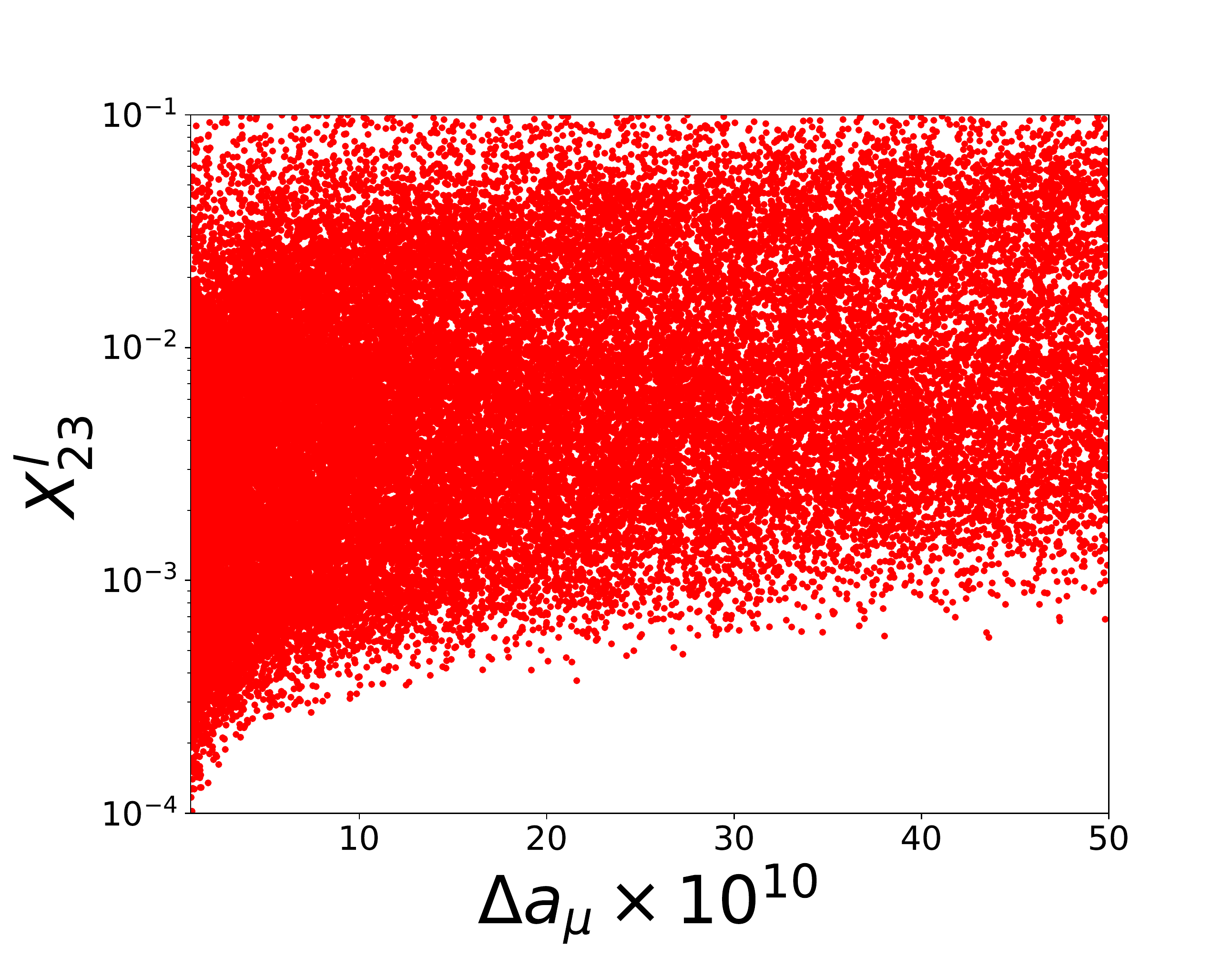}
\includegraphics[width=80mm]{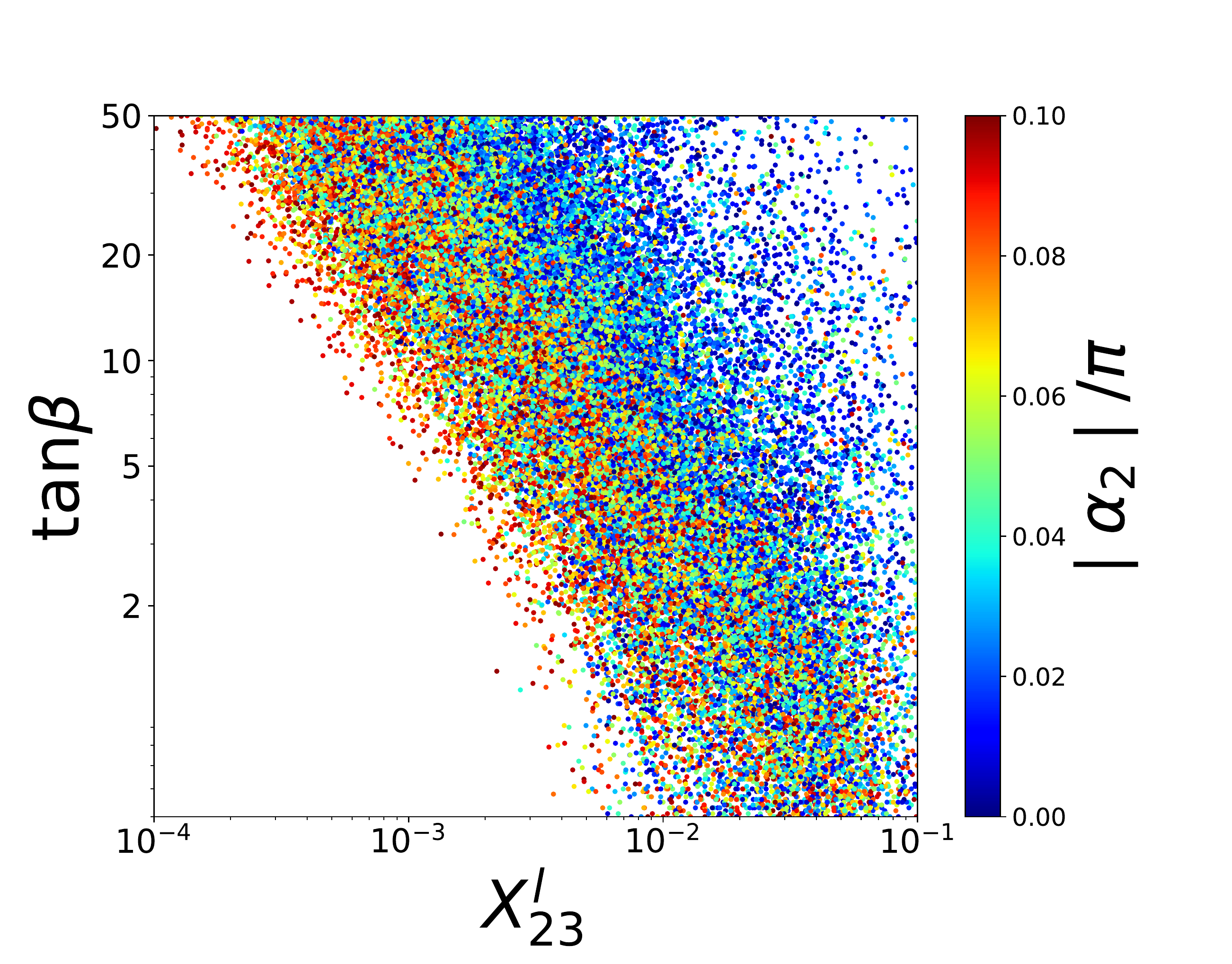}
\caption{The allowed parameter regions satisfying the muon $g-2$ value within $1 \times 10^{-10} < \Delta a_\mu < 50 \times 10^{-10}$, for $X^l_{32}=X^l_{23}$.}
\label{fig:muon g-2}
\end{figure}

In our analysis we consider (1) symmetric case $X_{23}^l = X_{32}^l$ and (2) anti-symmetric case $X_{23}^l = - X_{32}^l$ in estimating muon $g-2$ and LFV processes.
We then apply parameter range the same as shown in Eq.~\eqref{parameter range scalar} for scalar masses (except for lower limit of $\xi^0$ mass) and mixings, and $X_{ij}^\ell \in [10^{-5}, 1]$ for Yukawa couplings.
 The lower limit of $m_{\xi^0}$ is taken to be 10 GeV so that we can apply approximated form of LFV formulae~\footnote{Contribution to muon $g-2$ is not improved if we consider light $\xi^0$ case.}.

\noindent
{\bf (1) Symmetric case $X_{23}^l = X_{32}^l$}:

The left panel of Fig.~\ref{fig:muon g-2} shows the dependence of $X^l_{23}$ with $\Delta a_\mu$. We can get sizable contribution to muon $g-2$ for $X^l_{23} \gtrsim 10^{-4}$. On the right panel of the figure, the allowed region of $X^l_{23}$--$\tan\beta$ parameter space is shown with the value of parameter $\alpha_2$ indicated by color gradient. We can clearly see the correlation of $X^l_{23}$ with $\tan\beta$ and $\alpha_2$. For $X^l_{23} \lesssim 10^{-2}$, sizable muon $g-2$ can be obtained only for large $\tan\beta$ and large $\alpha_2$. This behavior can be understood by the fact that the dominant contribution to $Z_\phi$ in Eq.~\eqref{muon g-2 2} comes from the $\xi^0$ term and  small $X^l_{23}$ requires large $\tan\beta$ and large $\alpha_2$ since $R_{13}=\sin\alpha_2$. 
The contribution to $\Delta a_\mu$ from the pseudoscalar $A^0$ is negative in this case.
Thus it is preferred that $A^0$ is heavier than the other neutral scalar bosons so that negative contribution is relatively small~\footnote{Since we assume $m_{H^0} > m_{A^0}$ positive contribution from $H^0$ loop cannot overcome negative contribution from $A^0$. For the explanation of muon $g-2$ by $m_{H^0} < m_{A^0}$ case can be referred to ref.~\cite{Benbrik:2015evd}.}.

\begin{figure}[t!]
\includegraphics[width=80mm]{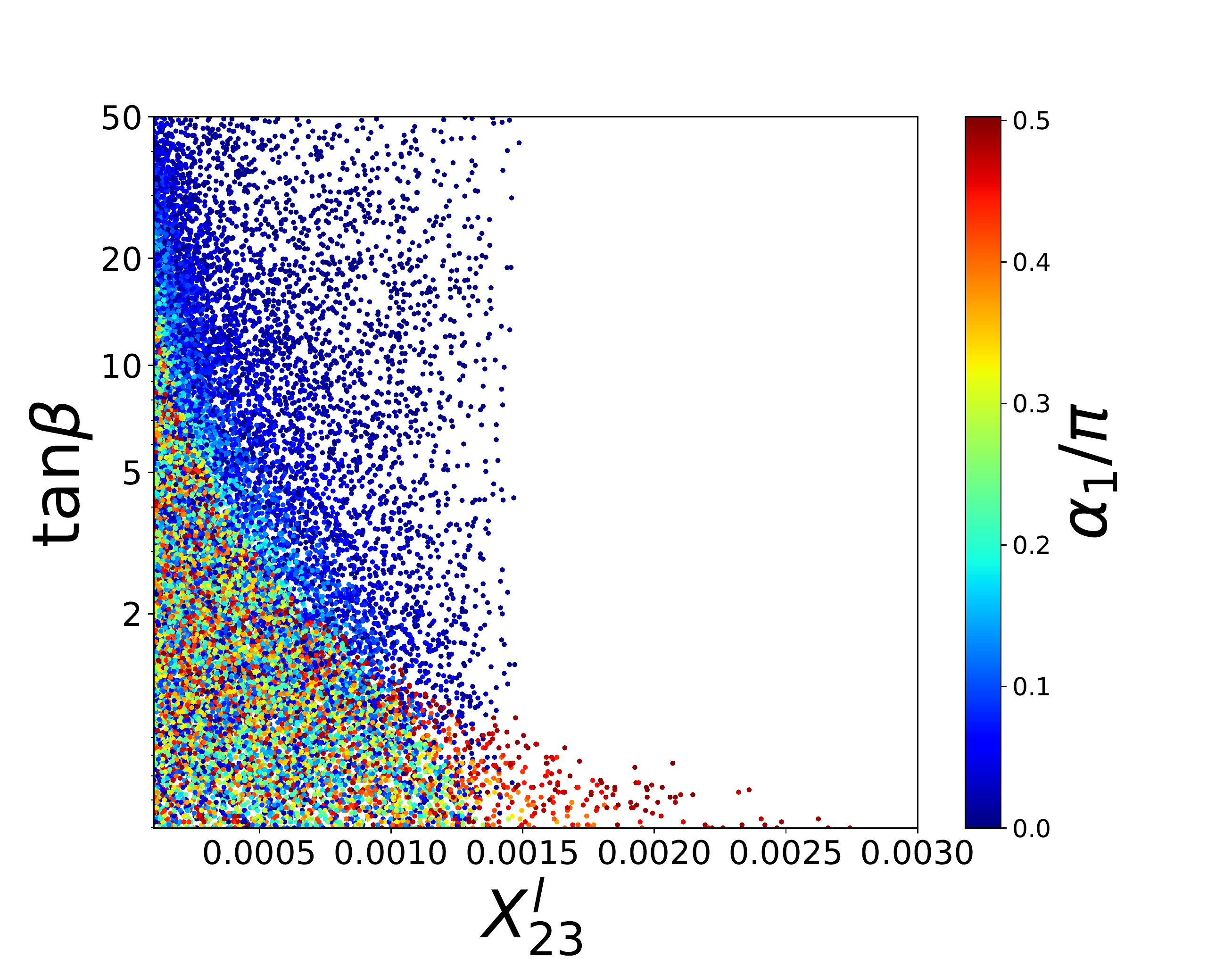}
\includegraphics[width=80mm]{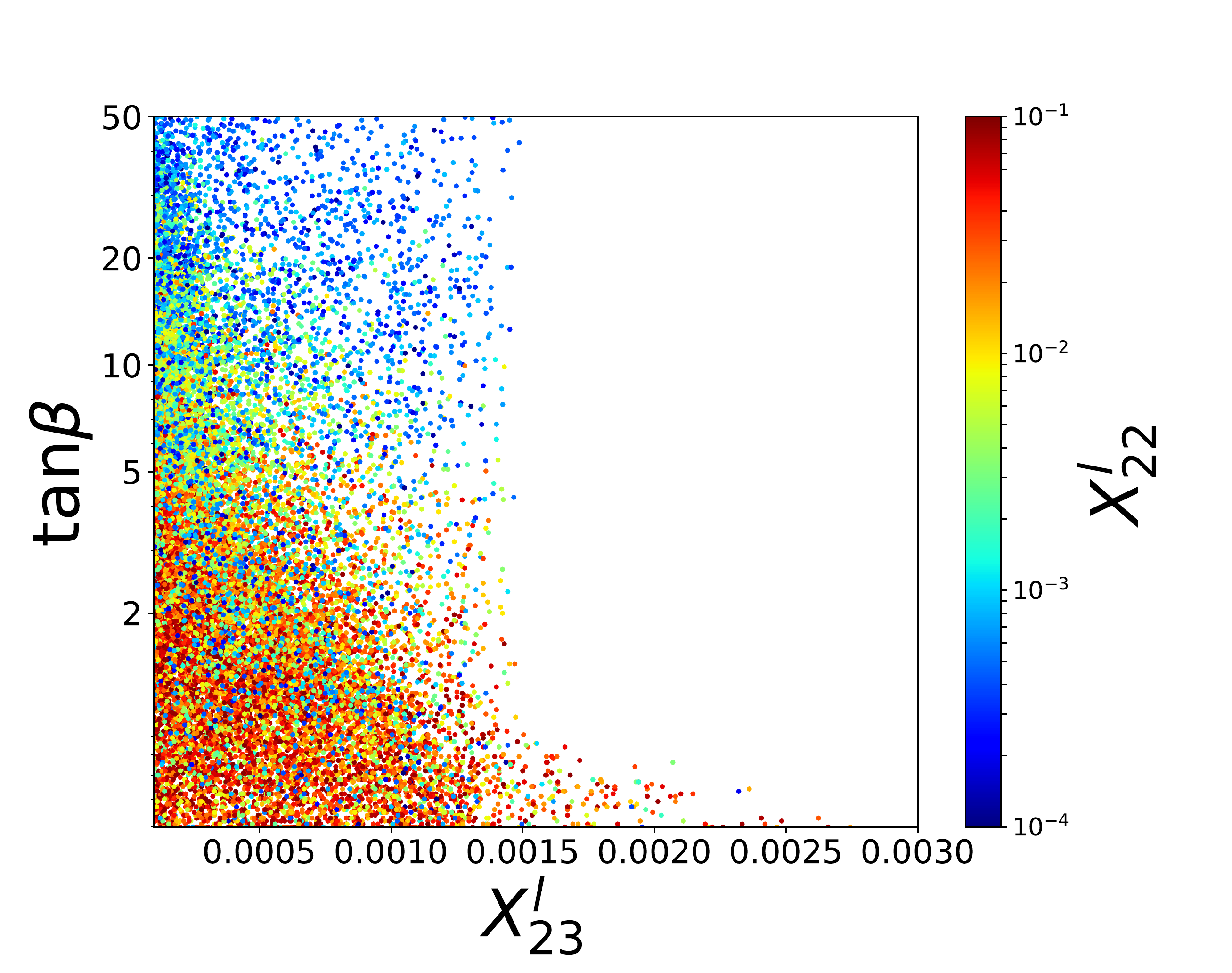}
\includegraphics[width=80mm]{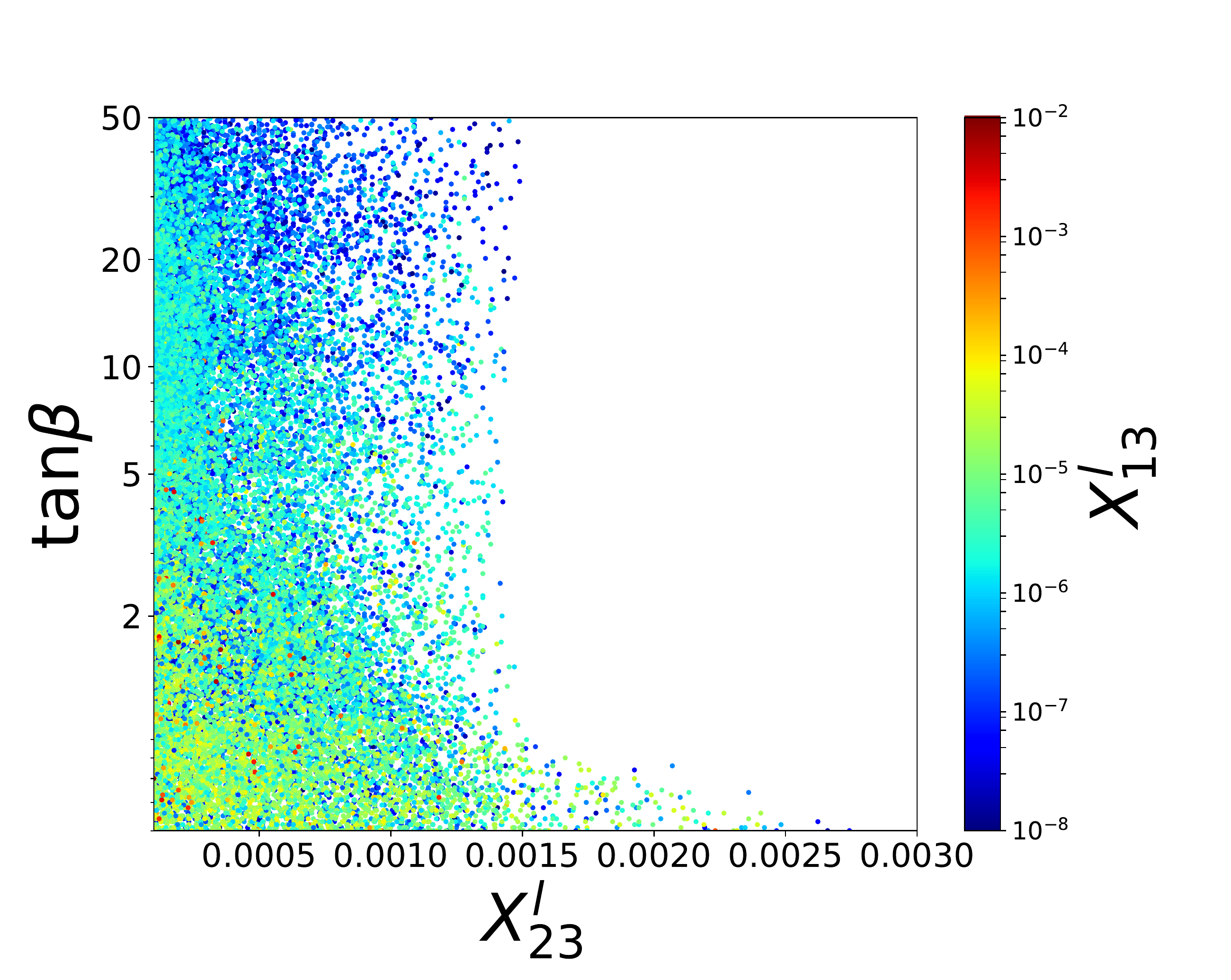}
\includegraphics[width=81mm]{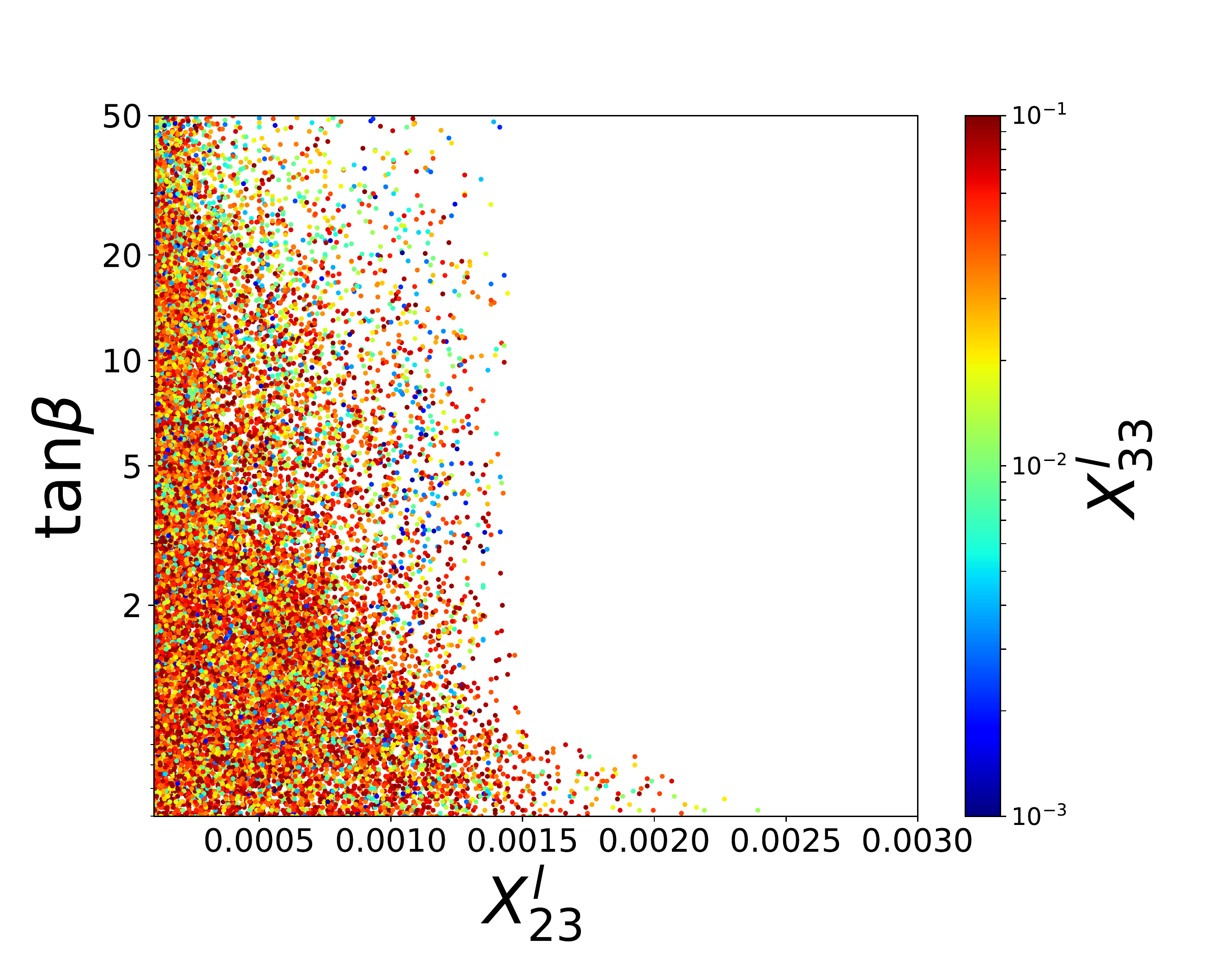}
\caption{The allowed parameter region satisfying constraints from $BR(h^0 \rightarrow \mu\tau)$, $BR(\tau\rightarrow 3\mu)$, $BR(\mu \rightarrow e\gamma)$ and $BR(\tau \rightarrow \mu\gamma)$.}
\label{fig:LFV}
\end{figure}

Fig.~\ref{fig:LFV} shows the allowed region of $X^l_{23}-\tan\beta$ space satisfying the LFV constraints associated with $BR(h^0 \rightarrow \mu\tau)$, $BR(\tau\rightarrow 3\mu)$, $BR(\mu \rightarrow e\gamma)$ and $BR(\tau \rightarrow \mu\gamma)$. The $BR(h^0\rightarrow\mu\tau)<0.25\%$ ~\cite{Sirunyan:2017xzt} puts a very strong constraint on the allowed range of $X^l_{23}$. In fact, the constraint of $BR(h^0 \rightarrow \mu \tau)$ restricts $X^l_{23}\lesssim 0.003$ for $\tan\beta <1$ and the upper limit of $X^l_{23}$ is $\sim 1.5\times 10^{-3}$ for $\tan\beta >1$ . Apart from that, there is a correlation between $\tan\beta$ and $\alpha_1$ which should be consistent with Fig.~\ref{fig:allowed1}. 
The same allowed region of $X^l_{23}-\tan\beta$ can be obtained for $X^l_{22}$ ranging from $(10^{-4}-10^{-1}$). The allowed region satisfying $BR(\mu \rightarrow e\gamma)$ constraints for $X^l_{13}$ in the range $(10^{-8}-10^{-2})$ as shown by the colorbar. For $X^l_{33}$ in the range $(10^{-3}-10^{-1})$, as shown in the colorbar, we can satisfy the  $BR(\mu \rightarrow e\gamma)$ constraint. 
Comparing Fig.~\ref{fig:muon g-2} and Fig.~\ref{fig:LFV}, there is a narrow strip of parameter space satisfying the LFV constraints and the muon $g-2$ constraint around $X^l_{23}\sim 10^{-4}-10^{-3}$. 

\begin{figure}[t!]
\includegraphics[width=78mm]{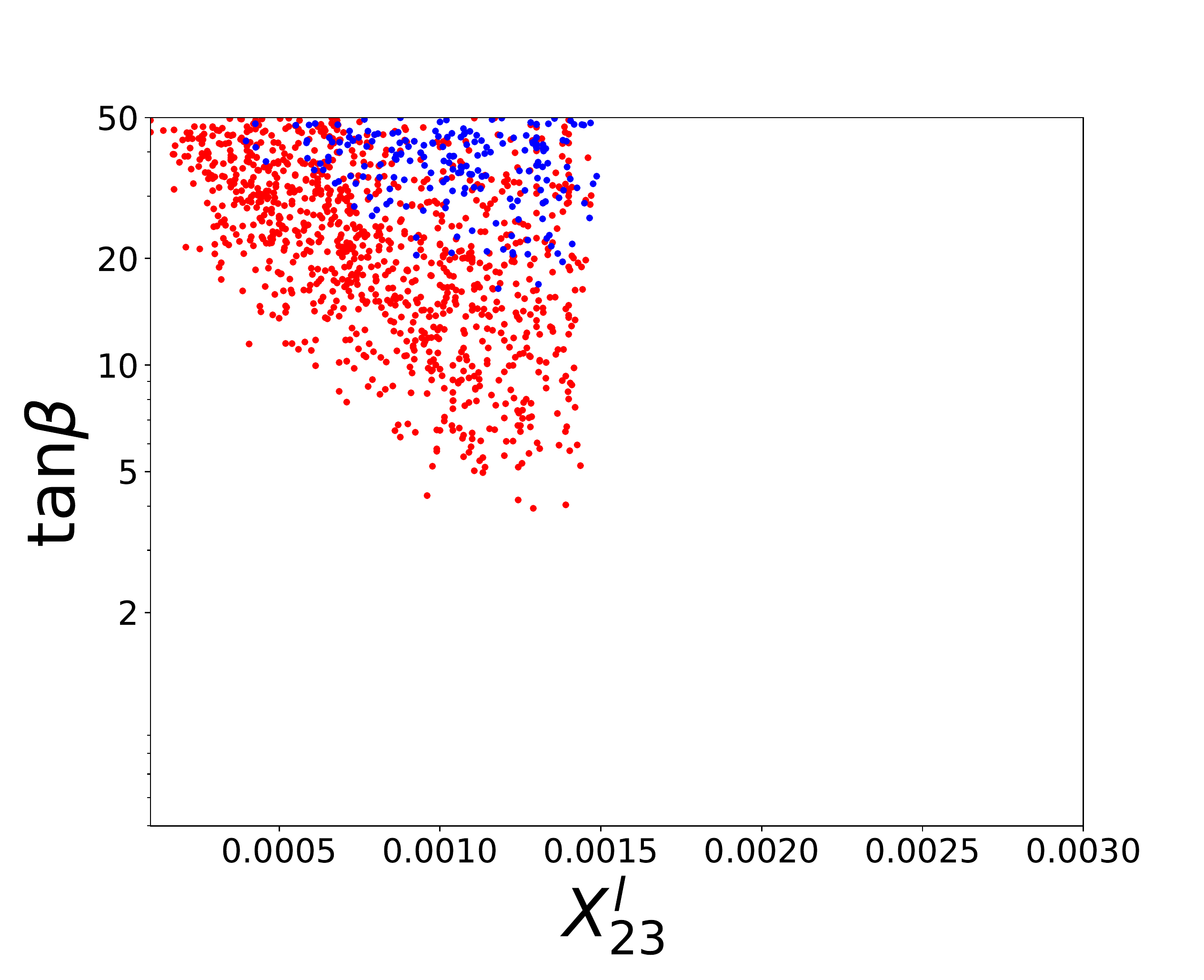}
\includegraphics[width=80mm]{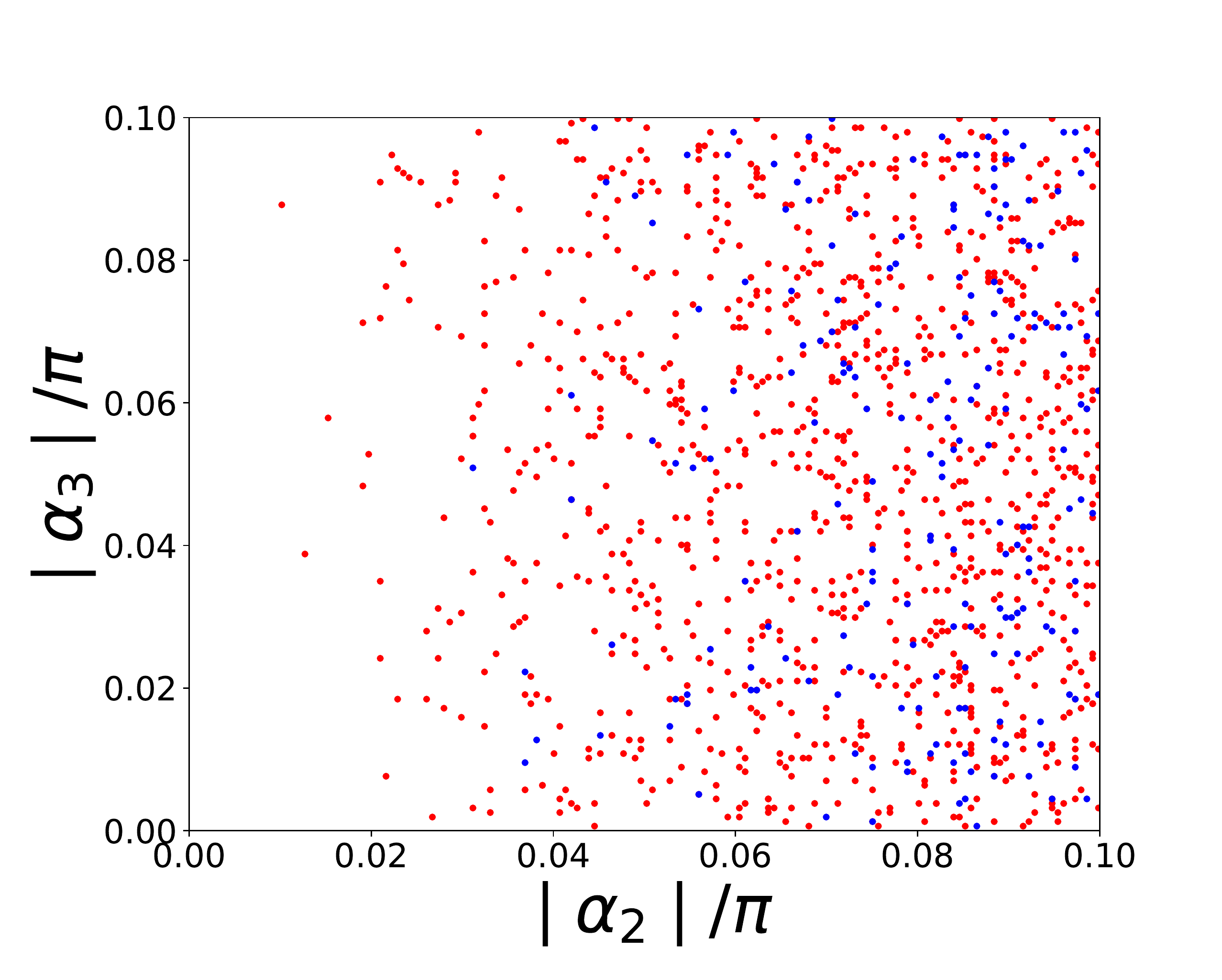}
\caption{The allowed parameter region satisfying the LFV constraints and muon $g-2$  value within $1 \times 10^{-10} < \Delta a_\mu < 50 \times 10^{-10}$ for $X^l_{32}=X^l_{23}$. Blue colored region realize muon $g-2$ within 2 $\sigma$ level.}
\label{fig:LFV muon magnetic moment} 
\end{figure}
Fig.~\ref{fig:LFV muon magnetic moment} shows the allowed region of parameter space consistent with the LFV constraints and muon $g-2$ constraint. For the allowed parameter space shown in the left panel, the $X^l_{23} - \tan\beta$ space is allowed for $X^l_{23}$ in the range $(10^{-4}- 1.5 \times 10^{-3})$ and $\tan\beta \gtrsim 3$. In the right pannel the allowed region in $\alpha_2 - \alpha_3$ space is shown where $\mid\alpha_2\mid/\pi \gtrsim 0.02$ is more dominant and is consistent with the right panel of Fig.~\ref{fig:muon g-2}.

\begin{figure}[t!]
\includegraphics[width=80mm]{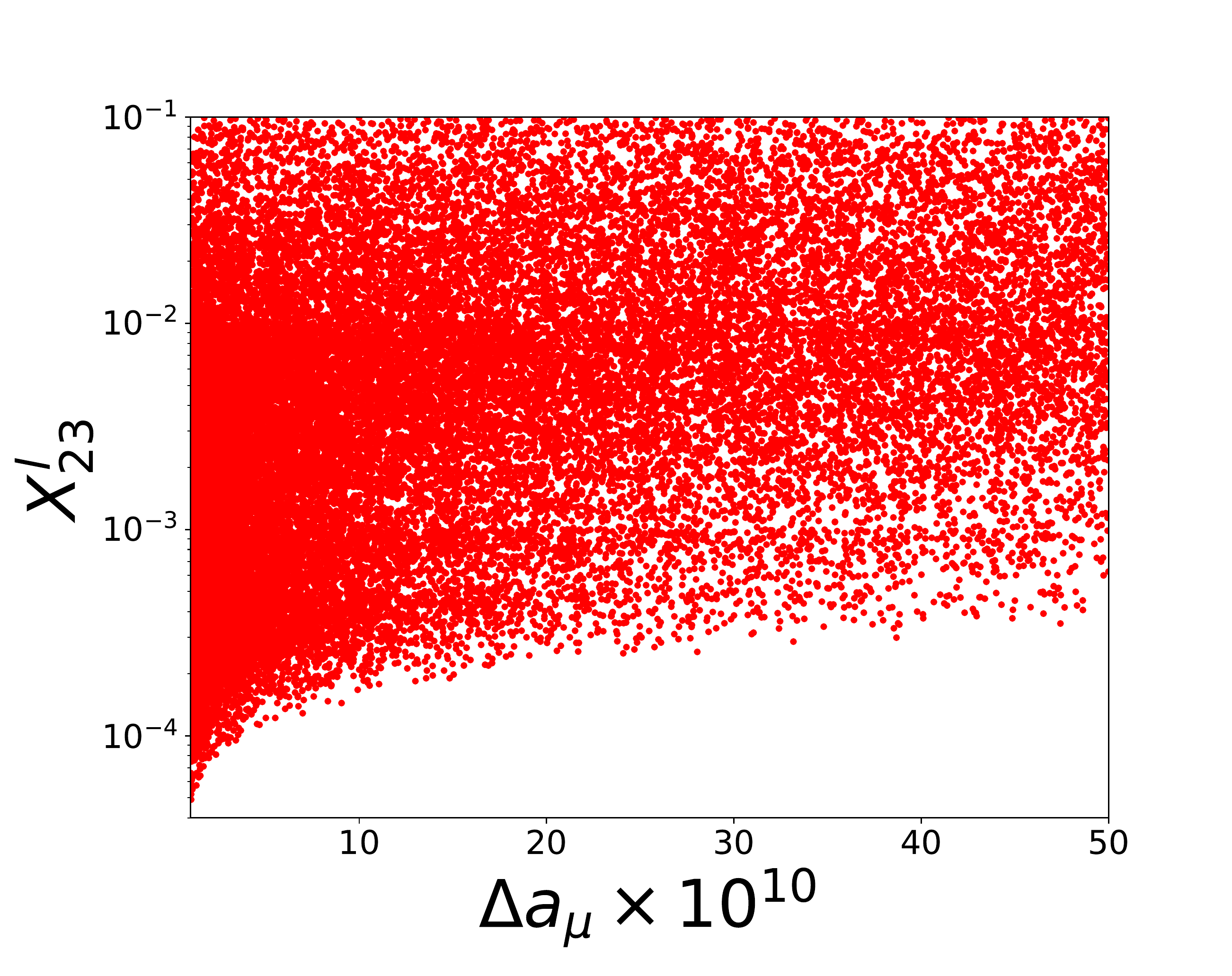}
\includegraphics[width=80mm]{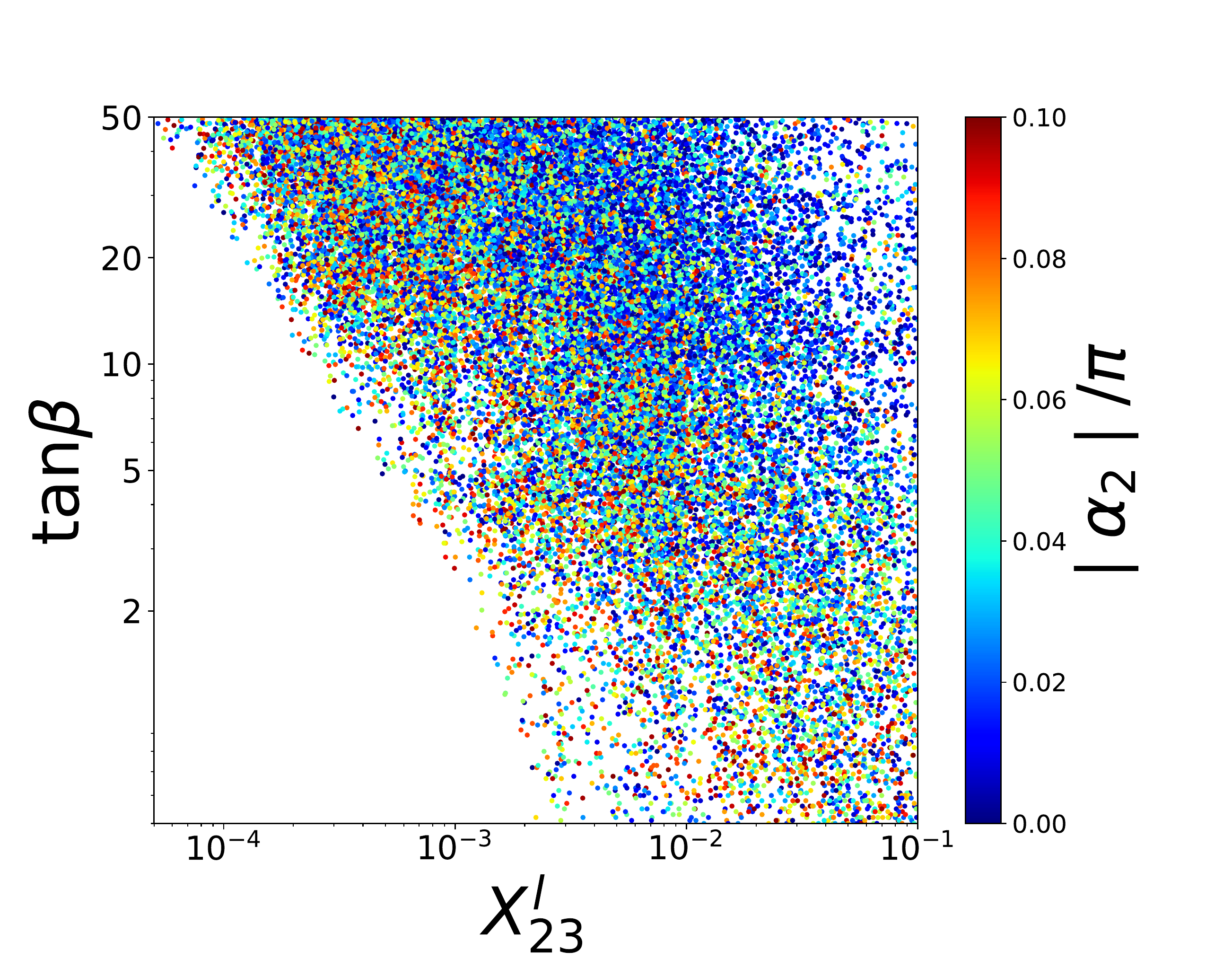}
\caption{The allowed parameter regions satisfying the muon $g-2$ value within $1 \times 10^{-10} < \Delta a_\mu < 50 \times 10^{-10}$, for $X^l_{32}=-X^l_{23}$.}
\label{fig:muon g-2 negative case}
\end{figure}

\noindent
{\bf (2) Anti-symmetric case $X^l_{32}=- X^l_{23}$}:
 
Here we estimate muon $g-2$ and LFV constraints for anti-symmetric Yukawa coupling case of $X^l_{32} = -X^l_{23}$ as in the symmetric case. 
We find that the parameter space can be relaxed in this case since the muon $g-2$ gets positive contribution from $A^0$ loop for the entire parameter space. 
In this antisymmetric situation the dominant contribution to $\Delta a_\mu$ comes from $A^0$, and hence $\alpha_2$ can be sufficiently low for $X^l_{23}\gtrsim 10^{-4}$ as shown in the right panel of Fig.~\ref{fig:muon g-2 negative case}. The left panel of Fig.~\ref{fig:muon g-2 negative case}  shows a clear improvement in the range of $X^l_{23}$ allowed by the muon $g-2$ constraint. We get sizable contribution for $X^l_{23}$ above $10^{-4}$.
We also find that the muon $g-2$ condition requires relatively large value for $\tan\beta$ for small  $X^l_{23}$ to enhance $A^0$ contribution. 
For LFV constraints we obtain the same behavior as the symmetric case shown in Fig.~\ref{fig:LFV} and parameter space explaining muon $g-2$ still can satisfiy the LFV constraints by proper choice of $X^l_{22}$, $X^l_{13}$ and $X^l_{33}$.

\begin{figure}[t!]
\includegraphics[width=78mm]{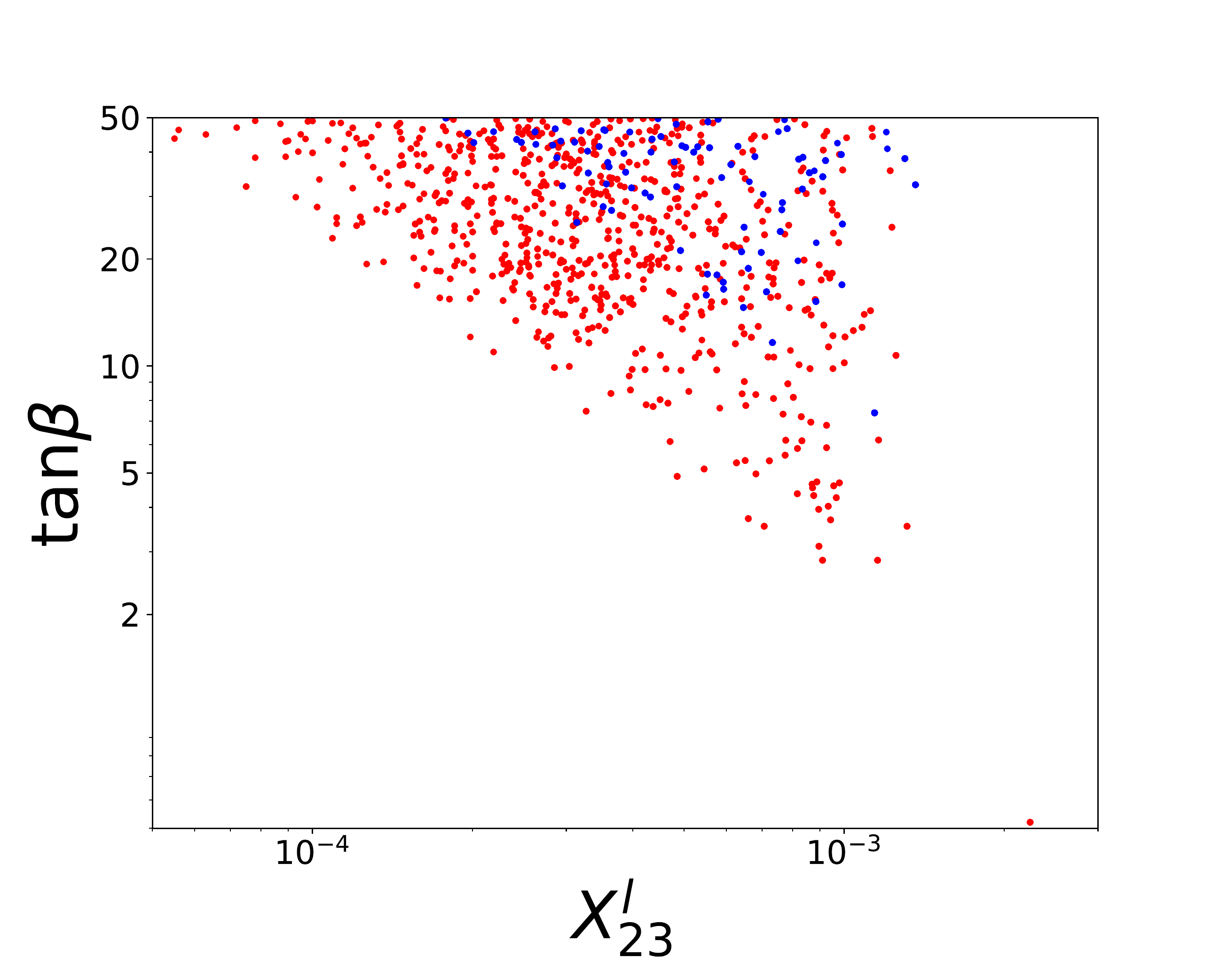}
\caption{The allowed parameter region satisfying the LFV constraints and muon $g-2$  value within $1 \times 10^{-10} < \Delta a_\mu < 50 \times 10^{-10}$ for $X^l_{32}=-X^l_{23}$. Blue colored region realize muon $g-2$ within 2 $\sigma$ level.}
\label{fig:LFV muon magnetic moment negative case} 
\end{figure}

In Fig.~\ref{fig:LFV muon magnetic moment negative case}, we show parameter space explaining muon $g-2$ and satisfying LFV constraints at the same time.
 Comparing the left panels of Fig.~\ref{fig:LFV muon magnetic moment} and Fig.~\ref{fig:LFV muon magnetic moment negative case}, we see that a larger region of $X^l_{23}-\tan\beta$ space is allowed for the antisymmetric case. 
 Finally, as mentioned above,  $\alpha_2$ can be sufficiently low since $A^0$ contribution is not suppressed by small $\alpha_2$.
 Thus parameter space explaining muon $g-2$ is qualitatively different in symmetric and anti-symmetric cases.

\section{Discussion: dark sector for realistic quark Yukawa}

\begin{figure}[t!]
\includegraphics[width=80mm]{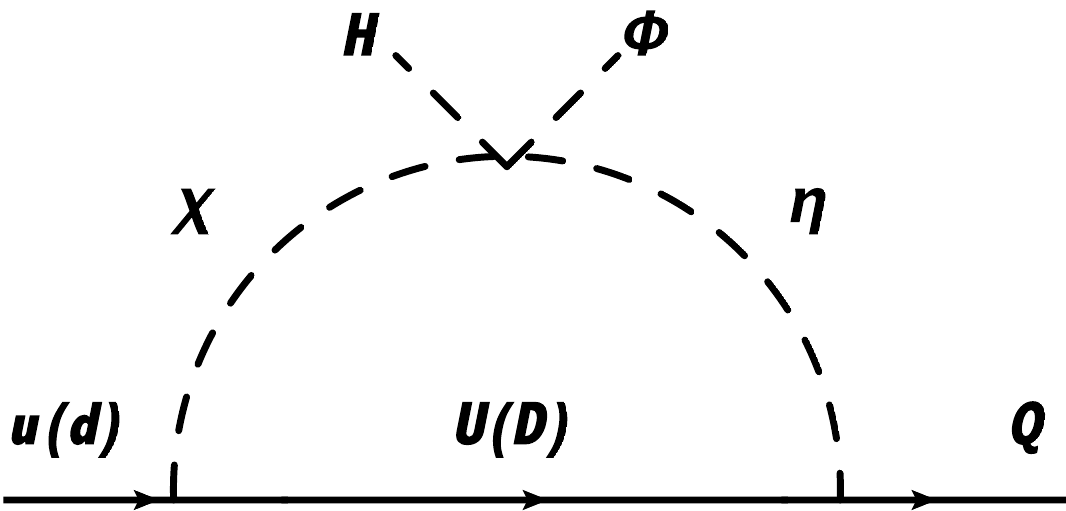}
\caption{One loop diagram generating quark Yukawa terms which is absent at tree level.}
\label{fig:diagram} 
\end{figure}

In this section, we briefly discuss possible dark sector to induce Yukawa terms in Eq.~\eqref{eff_Yukawa} which are absent at tree level due to our $U(1)_X$ charge assignment.
As a dark sector, for example, we introduce $Z_2$ odd fields as follows: vector like up-type quark $U$ which is $SU(2)$ singlet and $U(1)_X$ charge is $2 Q_H - 2 Q_{12}$,
vector like down-type quark $D$ which is $SU(2)$ singlet and $U(1)_X$ charge is $- Q_H + Q_{12}$, 
inert Higgs doublet $\eta$ which has $U(1)_X$ is $2 Q_H - 3 Q_{12}$ and inert singlet scalar $\chi$ without $U(1)_X$ charge.
Then we can write additional Lagrangian terms for quark Yukawa generation such that
\begin{align}
L_{\rm dark} = & f_{a} \bar Q_{L_a} \tilde \eta U_R + \tilde f \bar U_L u_{R_3} \chi + h_{} \bar Q_{L_3} \tilde \eta D_R + \tilde h_a \bar D_L d_{R_a} \chi
  + \lambda_D^{1} (H_{1}^\dagger \eta) \chi \phi +  \lambda_D^{2} (H_{2}^\dagger \eta) \chi \phi + h.c. ,
\end{align}
where $a = 1,2$ and we omit terms irrelevant to induce quark Yukawa.
From these interactions, we obtain the first two terms and the last two terms in Eq.~\eqref{eff_Yukawa} at one-loop level as shown in Fig.~\ref{fig:diagram}.
We can also extend dark sector and provide other terms of Eq.~\eqref{eff_Yukawa}.
In principle, we can tune free parameters to satisfy quark mixing while avoiding flavour constraints.
Interestingly dark sector includes dark matter candidates which are neutral component of $\eta$ and $\chi$.
In this paper we omit detailed analysis including dark matter physics and flavour constraints for dark sector since they are beyond our scope.

\section{Summary}

We have discussed a two Higgs doublet model with extra flavour depending $U(1)_X$ gauge symmetry where 
$Z'$ boson interactions can explain the Atomki anomaly by choosing appropriate charge assignment for the SM fermions.
We have shown $Z'$ coupling strengths with electron, neutron and proton which are consistent with the Atomki results.
In explaining the Atomki anomaly, we require $Z'$ boson mass around 17 MeV and gauge coupling of $\mathcal{O}(10^{-3})$
which indicates $U(1)_X$ breaking VEV of a SM singlet scalar to be $\mathcal{O}(10)$ GeV. 
As a result we have light scalar boson from the singlet scalar in this scenario. 

Then we have investigated scalar potential which contain two Higgs doublet and one singlet. 
Taking into account small VEV of the singlet, we have searched for allowed parameters such as scalar boson masses and mixings.
For scalar mixing, we have found stringent constraints come from the SM Higgs decay into $Z'Z'$ and light scalar pair $\xi^0 \xi^0$.

We also discussed muon $g-2$ and LFV constrains in our model.
The observed muon $g-2$ can be explained by flavour violating Yukawa couplings in general two Higgs doublet model 
due to $m_\tau$ enhancement from chiral flip inside an one loop diagram. 
We investigate constraints from LFV process associated with the flavour violating couplings to explain muon $g-2$.
In addition, effects of light scalar has been considered which provide changes from pure two Higgs doublet results.

Finally, we have discussed possible dark sector which can realize realistic quark mixing in our model 
where mixings associated with third generation quark is absent at renormalizable Lagrangian level 
due to our charge assignment for explaining the Atomki anomaly.
By introducing dark sector, we can generate such mixings at one loop level 
and realize observed CKM matrix in principle.
We have not discuss effect of dark sector in detail and it will be given in future works.

\section{Acknowledgements}
The work of T.N. was supported in part by KIAS Individual Grants, Grant No. PG054702 at Korea Institute for Advanced Study. The work of P.S. was supported by the Junior Research Group (JRG) Program at the Asia-Pacific Center for Theoretical
Physics (APCTP) through the Science and Technology Promotion Fund and Lottery Fund of the Korean Government and was supported by the Korean Local Governments-Gyeongsangbuk-do Province and Pohang City. The authors also acknowledge Harishyam Kumar for participating in the initial phase of the work. PS would like to thank Pankaj Jain for some useful discussions and comments.



\begin{thebibliography}{99}


\bibitem{Krasznahorkay:2015iga}
A.~J.~Krasznahorkay, M.~Csatl\'os, L.~Csige, Z.~G\'acsi, J.~Guly\'as, M.~Hunyadi, T.~J.~Ketel, A.~Krasznahorkay, I.~Kuti, B.~M.~Nyak\'o, L.~Stuhl, J.~Tim\'ar, T.~G.~Tornyi and Z.~Vajta,
Phys. Rev. Lett. \textbf{116} (2016) no.4, 042501
[arXiv:1504.01527 [nucl-ex]].

\bibitem{Krasznahorkay:2017qfd}
A.~J.~Krasznahorkay, M.~Csatl\'os, L.~Csige, J.~Guly\'as, M.~Hunyadi, T.~J.~Ketel, A.~Krasznahorkay, I.~Kuti, \'A.~Nagy, B.~M.~Nyak\'o, N.~Sas, J.~Tim\'ar and I.~Vajda,
EPJ Web Conf. \textbf{137} (2017), 08010

\bibitem{Krasznahorkay:2017gwn}
A.~J.~Krasznahorkay, M.~Csatl\'os, L.~Csige, J.~Guly\'as, T.~J.~Ketel, A.~Krasznahorkay, I.~Kuti, \'A.~Nagy, B.~M.~Nyak\'o, N.~Sas and J.~Tim\'ar,
EPJ Web Conf. \textbf{142} (2017), 01019

\bibitem{Krasznahorkay:2017bwh}
A.~J.~Krasznahorkay, M.~Csatl\'os, L.~Csige, J.~Guly\'as, M.~Hunyadi, T.~J.~Ketel, A.~Krasznahorkay, I.~Kuti, \'A.~Nagy, B.~M.~Nyak\'o, N.~Sas, J.~Tim\'ar and I.~Vajda,
PoS \textbf{BORMIO2017} (2017), 036

\bibitem{Krasznahorkay:2018snd}
A.~J.~Krasznahorkay, M.~Csatl\'os, L.~Csige, Z.~G\'acsi, J.~Guly\'as, \'A.~Nagy, N.~Sas, J.~Tim\'ar, T.~G.~Tornyi, I.~Vajda and A.~J.~Krasznahorkay,
J. Phys. Conf. Ser. \textbf{1056} (2018) no.1, 012028

\bibitem{Krasznahorkay:2019lyl}
A.~J.~Krasznahorkay, M.~Csatl\'os, L.~Csige, J.~Guly\'as, M.~Koszta, B.~Szihalmi, J.~Tim\'ar, D.~S.~Firak, \'A.~Nagy, N.~J.~Sas and G.~Cern,
[arXiv:1910.10459 [nucl-ex]].

\bibitem{Firak:2020eil}
D.~S.~Firak, A.~J.~Krasznahorkay, M.~Csatl\'os, L.~Csige, J.~Guly\'as, M.~Koszta, B.~Szihalmi, J.~Tim\'ar, \'A.~Nagy, N.~J.~Sas and A.~Krasznahorkay,
EPJ Web Conf. \textbf{232} (2020), 04005



\bibitem{Feng:2016jff}
J.~L.~Feng, B.~Fornal, I.~Galon, S.~Gardner, J.~Smolinsky, T.~M.~P.~Tait and P.~Tanedo,
Phys. Rev. Lett. \textbf{117} (2016) no.7, 071803
[arXiv:1604.07411 [hep-ph]].

\bibitem{Feng:2016ysn}
J.~L.~Feng, B.~Fornal, I.~Galon, S.~Gardner, J.~Smolinsky, T.~M.~P.~Tait and P.~Tanedo,
Phys. Rev. D \textbf{95} (2017) no.3, 035017
[arXiv:1608.03591 [hep-ph]].

\bibitem{Seto:2016pks}
O.~Seto and T.~Shimomura,
Phys. Rev. D \textbf{95} (2017) no.9, 095032
[arXiv:1610.08112 [hep-ph]].

\bibitem{Gu:2016ege}
P.~H.~Gu and X.~G.~He,
Nucl. Phys. B \textbf{919} (2017), 209-217
[arXiv:1606.05171 [hep-ph]].



\bibitem{Neves:2017rcn}
M.~J.~Neves and E.~M.~C.~Abreu,
Acta Phys. Polon. B \textbf{51} (2020), 909
[arXiv:1704.02491 [hep-ph]].

\bibitem{DelleRose:2018eic}
L.~Delle Rose, S.~Khalil, S.~J.~D.~King, S.~Moretti and A.~M.~Thabt,
Phys. Rev. D \textbf{99} (2019) no.5, 055022
[arXiv:1811.07953 [hep-ph]].

\bibitem{Pulice:2019xel}
B.~Puli\c{c}e,
[arXiv:1911.10482 [hep-ph]].

\bibitem{DelleRose:2017xil}
L.~Delle Rose, S.~Khalil and S.~Moretti,
Phys. Rev. D \textbf{96} (2017) no.11, 115024
[arXiv:1704.03436 [hep-ph]].

\bibitem{Kitahara:2016zyb}
T.~Kitahara and Y.~Yamamoto,
Phys. Rev. D \textbf{95} (2017) no.1, 015008
[arXiv:1609.01605 [hep-ph]].

\bibitem{Jia:2017iyc}
L.~B.~Jia,
Eur. Phys. J. C \textbf{78} (2018) no.2, 112
[arXiv:1710.03906 [hep-ph]].

\bibitem{Jia:2018mkc}
L.~B.~Jia, X.~J.~Deng and C.~F.~Liu,
Eur. Phys. J. C \textbf{78} (2018) no.11, 956
[arXiv:1809.00177 [hep-ph]].

%

\bibitem{Neves:2018bay}
M.~J.~Neves, L.~Labre, L.~S.~Miranda and E.~Abreu, M.C.,
Int. J. Mod. Phys. A \textbf{33} (2018) no.25, 1850148
[arXiv:1802.10449 [hep-ph]].

\bibitem{BORDES:2019wcp}
J.~Bordes, H.~M.~Chan and S.~T.~Tsou,
Int. J. Mod. Phys. A \textbf{34} (2019) no.25, 1950140
[arXiv:1906.09229 [hep-ph]].

\bibitem{Nam:2019osu}
C.~H.~Nam,
Eur. Phys. J. C \textbf{80} (2020) no.3, 231
[arXiv:1907.09819 [hep-ph]].

\bibitem{Krasnikov:2019dgh}
N.~V.~Krasnikov,
Mod. Phys. Lett. A \textbf{35} (2020) no.15, 2050116
[arXiv:1912.11689 [hep-ph]].


\bibitem{Wong:2020hjc}
C.~Y.~Wong,
[arXiv:2001.04864 [nucl-th]].

\bibitem{Tursunov:2020wfy}
E.~M.~Tursunov,
[arXiv:2001.08995 [nucl-th]].

\bibitem{Kirpichnikov:2020tcf}
D.~V.~Kirpichnikov, V.~E.~Lyubovitskij and A.~S.~Zhevlakov,
[arXiv:2002.07496 [hep-ph]].

\bibitem{Hati:2020fzp}
C.~Hati, J.~Kriewald, J.~Orloff and A.~M.~Teixeira,
[arXiv:2005.00028 [hep-ph]].

\bibitem{Chen:2020arr}
H.~X.~Chen,
[arXiv:2006.01018 [hep-ph]].

\bibitem{Zhang:2020ukq}
X.~Zhang and G.~A.~Miller,
[arXiv:2008.11288 [hep-ph]].

\bibitem{Feng:2020mbt}
J.~L.~Feng, T.~M.~P.~Tait and C.~B.~Verhaaren,
Phys. Rev. D \textbf{102} (2020) no.3, 036016
[arXiv:2006.01151 [hep-ph]].

\bibitem{Seto:2020jal}
O.~Seto and T.~Shimomura,
[arXiv:2006.05497 [hep-ph]].



   \bibitem{PDG} M. Tanabashi et al. (Particle Data Group), Phys. Rev. D {\bf 98}, 030001 (2018).
   
   
\bibitem{Keshavarzi:2018mgv} 
  A.~Keshavarzi, D.~Nomura and T.~Teubner,
  Phys.\ Rev.\ D {\bf 97}, no. 11, 114025 (2018)
  [arXiv:1802.02995 [hep-ph]].
  
  
 
\bibitem{Davier:2010nc}
M.~Davier, A.~Hoecker, B.~Malaescu and Z.~Zhang,
Eur. Phys. J. C \textbf{71} (2011), 1515
[erratum: Eur. Phys. J. C \textbf{72} (2012), 1874]
[arXiv:1010.4180 [hep-ph]].
 
\bibitem{Davier:2017zfy}
M.~Davier, A.~Hoecker, B.~Malaescu and Z.~Zhang,
Eur. Phys. J. C \textbf{77} (2017) no.12, 827
[arXiv:1706.09436 [hep-ph]].
 
  
\bibitem{Davier:2019can}
M.~Davier, A.~Hoecker, B.~Malaescu and Z.~Zhang,
Eur. Phys. J. C \textbf{80} (2020) no.3, 241
[erratum: Eur. Phys. J. C \textbf{80} (2020) no.5, 410]
[arXiv:1908.00921 [hep-ph]].
  
  
\bibitem{Aoyama:2020ynm}
T.~Aoyama, N.~Asmussen, M.~Benayoun, J.~Bijnens, T.~Blum, M.~Bruno, I.~Caprini, C.~M.~Carloni Calame, M.~C\`e and G.~Colangelo, \textit{et al.}
[arXiv:2006.04822 [hep-ph]].

  
  

\bibitem{Grange:2015fou} 
  J.~Grange {\it et al.} [Muon g-2 Collaboration],
  arXiv:1501.06858 [physics.ins-det].
  

  
\bibitem{Otani:2015jra} 
  M.~Otani [E34 Collaboration],
  JPS Conf.\ Proc.\  {\bf 8}, 025008 (2015).

\bibitem{Hong:2018kqx} 
  R.~Hong [Muon g-2 Collaboration],
  arXiv:1810.03729 [physics.ins-det].
  
\bibitem{Borsanyi:2020mff} 
  S.~Borsanyi {\it et al.},
  arXiv:2002.12347 [hep-lat].

  
\bibitem{Crivellin:2020zul} 
  A.~Crivellin, M.~Hoferichter, C.~A.~Manzari and M.~Montull,
  arXiv:2003.04886 [hep-ph].
  
\bibitem{deRafael:2020uif}
E.~de Rafael,
Phys. Rev. D \textbf{102} (2020) no.5, 056025
[arXiv:2006.13880 [hep-ph]].
  
  
\bibitem{Passera:2008jk}
M.~Passera, W.~J.~Marciano and A.~Sirlin,
Phys. Rev. D \textbf{78} (2008), 013009
[arXiv:0804.1142 [hep-ph]].


  

\bibitem{Czarnecki:2001pv} 
  A.~Czarnecki and W.~J.~Marciano,
  Phys.\ Rev.\ D {\bf 64}, 013014 (2001)
  [hep-ph/0102122].
  
\bibitem{Gninenko:2001hx} 
  S.~N.~Gninenko and N.~V.~Krasnikov,
  Phys.\ Lett.\ B {\bf 513}, 119 (2001)
  [hep-ph/0102222].
  
\bibitem{Ma:2001mr} 
  E.~Ma and M.~Raidal,
  Phys.\ Rev.\ Lett.\  {\bf 87}, 011802 (2001)
  Erratum: [Phys.\ Rev.\ Lett.\  {\bf 87}, 159901 (2001)]
  [hep-ph/0102255].
  
\bibitem{Chen:2001kn} 
  C.~H.~Chen and C.~Q.~Geng,
  Phys.\ Lett.\ B {\bf 511}, 77 (2001)
  [hep-ph/0104151].
  
\bibitem{Ma:2001md} 
  E.~Ma, D.~P.~Roy and S.~Roy,
  Phys.\ Lett.\ B {\bf 525}, 101 (2002)
  [hep-ph/0110146].
  
\bibitem{Benbrik:2015evd} 
  R.~Benbrik, C.~H.~Chen and T.~Nomura,
  Phys.\ Rev.\ D {\bf 93}, no. 9, 095004 (2016)
  [arXiv:1511.08544 [hep-ph]].
  
\bibitem{Nomura:2016rjf} 
  T.~Nomura and H.~Okada,
  Phys.\ Lett.\ B {\bf 756}, 295 (2016)
  [arXiv:1601.07339 [hep-ph]].
  
\bibitem{Baek:2016kud} 
  S.~Baek, T.~Nomura and H.~Okada,
  Phys.\ Lett.\ B {\bf 759}, 91 (2016)
  [arXiv:1604.03738 [hep-ph]].
  
  
      
\bibitem{Altmannshofer:2016oaq} 
  W.~Altmannshofer, M.~Carena and A.~Crivellin,
  Phys.\ Rev.\ D {\bf 94}, no. 9, 095026 (2016)
  [arXiv:1604.08221 [hep-ph]].
  
\bibitem{Chen:2016dip} 
  C.~H.~Chen, T.~Nomura and H.~Okada,
  Phys.\ Rev.\ D {\bf 94}, no. 11, 115005 (2016)
  [arXiv:1607.04857 [hep-ph]].
  
\bibitem{Lindner:2016bgg}
M.~Lindner, M.~Platscher and F.~S.~Queiroz,
Phys. Rept. \textbf{731} (2018), 1-82
[arXiv:1610.06587 [hep-ph]].


  
  
\bibitem{Megias:2017dzd}
E.~Megias, M.~Quiros and L.~Salas,
JHEP \textbf{05} (2017), 016
[arXiv:1701.05072 [hep-ph]].

  
  
\bibitem{Lee:2017ekw} 
  S.~Lee, T.~Nomura and H.~Okada,
  Nucl.\ Phys.\ B {\bf 931}, 179 (2018)
  [arXiv:1702.03733 [hep-ph]].
 
\bibitem{Chen:2017hir} 
  C.~H.~Chen, T.~Nomura and H.~Okada,
  Phys.\ Lett.\ B {\bf 774}, 456 (2017)
  [arXiv:1703.03251 [hep-ph]].
  
\bibitem{Das:2017ski} 
  A.~Das, T.~Nomura, H.~Okada and S.~Roy,
  Phys.\ Rev.\ D {\bf 96}, no. 7, 075001 (2017)
  [arXiv:1704.02078 [hep-ph]].
  
\bibitem{Kowalska:2017iqv} 
  K.~Kowalska and E.~M.~Sessolo,
  JHEP {\bf 1709}, 112 (2017)
  [arXiv:1707.00753 [hep-ph]].
  
\bibitem{Terazawa:2018pdc}
H.~Terazawa,
Nonlin. Phenom. Complex Syst. \textbf{21} (2018) no.3, 268-272
  
\bibitem{Calibbi:2018rzv} 
  L.~Calibbi, R.~Ziegler and J.~Zupan,
  JHEP {\bf 1807}, 046 (2018)
  [arXiv:1804.00009 [hep-ph]].
  
\bibitem{Barman:2018jhz} 
  B.~Barman, D.~Borah, L.~Mukherjee and S.~Nandi,
  Phys.\ Rev.\ D {\bf 100}, no. 11, 115010 (2019)
  [arXiv:1808.06639 [hep-ph]].
  
  
\bibitem{Liu:2018xkx}
J.~Liu, C.~E.~M.~Wagner and X.~P.~Wang,
JHEP \textbf{03} (2019), 008
[arXiv:1810.11028 [hep-ph]].
  
  
  
\bibitem{Nomura:2019btk}
T.~Nomura and H.~Okada,
Phys. Rev. D \textbf{101} (2020) no.1, 015021
[arXiv:1903.05958 [hep-ph]].

\bibitem{Nomura:2019wlo}
T.~Nomura and P.~Sanyal,
Phys. Rev. D \textbf{100} (2019) no.11, 115036
[arXiv:1907.02718 [hep-ph]].

\bibitem{Liu:2020qgx}
J.~Liu, N.~McGinnis, C.~E.~M.~Wagner and X.~P.~Wang,
JHEP \textbf{04} (2020), 197
[arXiv:2001.06522 [hep-ph]].

\bibitem{Kumar:2020web}
N.~Kumar, T.~Nomura and H.~Okada,
[arXiv:2002.12218 [hep-ph]].


 
\bibitem{Chen:2020jvl}
C.~H.~Chen and T.~Nomura,
[arXiv:2003.07638 [hep-ph]].

\bibitem{deJesus:2020upp}
A.~S.~De Jesus, S.~Kovalenko, F.~S.~Queiroz, C.~Siqueira and K.~Sinha,
Phys. Rev. D \textbf{102} (2020) no.3, 035004
[arXiv:2004.01200 [hep-ph]].


 
  
\bibitem{Das:2016zue}
A.~Das, S.~Oda, N.~Okada and D.~s.~Takahashi,
Phys. Rev. D \textbf{93} (2016) no.11, 115038
[arXiv:1605.01157 [hep-ph]].
  
\bibitem{Das:2017flq}
A.~Das, N.~Okada and D.~Raut,
Phys. Rev. D \textbf{97} (2018) no.11, 115023
[arXiv:1710.03377 [hep-ph]].
  
\bibitem{Das:2017deo}
A.~Das, N.~Okada and D.~Raut,
Eur. Phys. J. C \textbf{78} (2018) no.9, 696
[arXiv:1711.09896 [hep-ph]].
  
\bibitem{Das:2018tbd}
A.~Das, N.~Okada, S.~Okada and D.~Raut,
Phys. Lett. B \textbf{797} (2019), 134849
[arXiv:1812.11931 [hep-ph]].
  
\bibitem{Das:2019pua}
A.~Das, S.~Goswami, K.~N.~Vishnudath and T.~Nomura,
Phys. Rev. D \textbf{101} (2020) no.5, 055026
[arXiv:1905.00201 [hep-ph]].
  
\bibitem{Das:2019fee}
A.~Das, P.~S.~B.~Dev and N.~Okada,
Phys. Lett. B \textbf{799} (2019), 135052
[arXiv:1906.04132 [hep-ph]].
  
\bibitem{Chiang:2019ajm}
C.~W.~Chiang, G.~Cottin, A.~Das and S.~Mandal,
JHEP \textbf{12} (2019), 070
[arXiv:1908.09838 [hep-ph]].

\bibitem{Nomura:2020azp}
T.~Nomura, H.~Okada and Y.~Uesaka,
[arXiv:2005.05527 [hep-ph]].

  
\bibitem{Nomura:2020dzw}
T.~Nomura, H.~Okada and Y.~Uesaka,
[arXiv:2008.02673 [hep-ph]].

  


\bibitem{Becirevic:2016zri}
D.~Becirevic, O.~Sumensari and R.~Zukanovich Funchal,
Eur. Phys. J. C \textbf{76} (2016) no.3, 134
[arXiv:1602.00881 [hep-ph]].

\bibitem{Kumar:2019qbv}
J.~Kumar and D.~London,
Phys. Rev. D \textbf{99} (2019) no.7, 073008
[arXiv:1901.04516 [hep-ph]].

\bibitem{Dror:2017nsg}
J.~A.~Dror, R.~Lasenby and M.~Pospelov,
Phys. Rev. D \textbf{96} (2017) no.7, 075036
[arXiv:1707.01503 [hep-ph]].






\bibitem{Muhlleitner:2016mzt}
M.~Muhlleitner, M.~O.~P.~Sampaio, R.~Santos and J.~Wittbrodt,
JHEP \textbf{03} (2017), 094
[arXiv:1612.01309 [hep-ph]].

\bibitem{Bian:2017xzg} 
  L.~Bian, H.~M.~Lee and C.~B.~Park,
  Eur.\ Phys.\ J.\ C {\bf 78}, no. 4, 306 (2018)
  [arXiv:1711.08930 [hep-ph]].
  


\bibitem{Banerjee:2019hmi}
D.~Banerjee \textit{et al.} [NA64],
Phys. Rev. D \textbf{101} (2020) no.7, 071101
[arXiv:1912.11389 [hep-ex]].


\bibitem{Langacker:2008yv}
P.~Langacker,
Rev. Mod. Phys. \textbf{81} (2009), 1199-1228
[arXiv:0801.1345 [hep-ph]].


\bibitem{Celis:2013ixa}
A.~Celis, V.~Ilisie and A.~Pich,
JHEP \textbf{12} (2013), 095
[arXiv:1310.7941 [hep-ph]].

\bibitem{Dumont:2014wha}
B.~Dumont, J.~F.~Gunion, Y.~Jiang and S.~Kraml,
Phys. Rev. D \textbf{90} (2014), 035021
[arXiv:1405.3584 [hep-ph]].

\bibitem{Bernon:2014nxa}
J.~Bernon, J.~F.~Gunion, Y.~Jiang and S.~Kraml,
Phys. Rev. D \textbf{91} (2015) no.7, 075019
[arXiv:1412.3385 [hep-ph]].

\bibitem{Craig:2015jba}
N.~Craig, F.~D'Eramo, P.~Draper, S.~Thomas and H.~Zhang,
JHEP \textbf{06} (2015), 137
[arXiv:1504.04630 [hep-ph]].

\bibitem{Bernon:2015qea}
J.~Bernon, J.~F.~Gunion, H.~E.~Haber, Y.~Jiang and S.~Kraml,
Phys. Rev. D \textbf{92} (2015) no.7, 075004
[arXiv:1507.00933 [hep-ph]].

\bibitem{Arhrib:2015maa}
A.~Arhrib, R.~Benbrik, C.~H.~Chen, M.~Gomez-Bock and S.~Semlali,
Eur. Phys. J. C \textbf{76} (2016) no.6, 328
[arXiv:1508.06490 [hep-ph]].

\bibitem{Keus:2015hva}
V.~Keus, S.~F.~King, S.~Moretti and K.~Yagyu,
JHEP \textbf{04} (2016), 048
[arXiv:1510.04028 [hep-ph]].

\bibitem{Aggleton:2016tdd}
R.~Aggleton, D.~Barducci, N.~E.~Bomark, S.~Moretti and C.~Shepherd-Themistocleous,
JHEP \textbf{02} (2017), 035
[arXiv:1609.06089 [hep-ph]].

\bibitem{Chen:2018uim}
N.~Chen, C.~Du, Y.~Wu and X.~J.~Xu,
Phys. Rev. D \textbf{99} (2019) no.3, 035011
[arXiv:1810.04689 [hep-ph]].

\bibitem{Sanyal:2019xcp}
P.~Sanyal,
Eur. Phys. J. C \textbf{79} (2019) no.11, 913
[arXiv:1906.02520 [hep-ph]].


\bibitem{Kling:2020hmi}
F.~Kling, S.~Su and W.~Su,
JHEP \textbf{06} (2020), 163
[arXiv:2004.04172 [hep-ph]].

\bibitem{Aiko:2020ksl}
M.~Aiko, S.~Kanemura, M.~Kikuchi, K.~Mawatari, K.~Sakurai and K.~Yagyu,
[arXiv:2010.15057 [hep-ph]].


\bibitem{Bechtle:2013xfa}
P.~Bechtle, S.~Heinemeyer, O.~St\r{a}l, T.~Stefaniak and G.~Weiglein,
Eur. Phys. J. C \textbf{74} (2014) no.2, 2711
doi:10.1140/epjc/s10052-013-2711-4
[arXiv:1305.1933 [hep-ph]].


\bibitem{ATLAS:2018doi} 
  The ATLAS collaboration [ATLAS Collaboration],
  ATLAS-CONF-2018-031.
  
\bibitem{Sirunyan:2018koj} 
  A.~M.~Sirunyan {\it et al.} [CMS Collaboration],
  [arXiv:1809.10733 [hep-ex]].
  
\bibitem{Sirunyan:2017xzt}
A.~M.~Sirunyan \textit{et al.} [CMS],
JHEP \textbf{06} (2018), 001
[arXiv:1712.07173 [hep-ex]].




\end{thebibliography}
\end{document}